\documentclass{emulateapj}
\usepackage{apjfonts}
\usepackage{graphicx}

\usepackage{amsmath}

\shorttitle{Direct Numerical Simulations of Cosmic-ray Acceleration}
\shortauthors{T. INOUE et al.}

\begin{document}

\title{
Direct Numerical Simulations of Cosmic-ray Acceleration at Dense Circumstellar Medium: Magnetic Field Amplification by Bell Instability and Maximum Energy}
\author{Tsuyoshi Inoue\altaffilmark{1}, Alexandre Marcowith\altaffilmark{2}, Gwenael Giacinti\altaffilmark{3}, Allard Jan van Marle\altaffilmark{2}, and Shogo Nishino\altaffilmark{1}}
\altaffiltext{1}{Department of Physics, Graduate School of Science, Nagoya University, Furo-cho, Chikusa-ku, Nagoya 464-8602, Japan; tsuyoshi.inoue@nagoya-u.jp}
\altaffiltext{2}{Laboratoire Universe et Particules de Montpellier (LUPM) Universit\'{e} Montpellier, CNRS/IN2P3, CC72, place Eug\`{e}ne Bataillon, 34095, Montpellier Cedex 5, France}
\altaffiltext{3}{Max-Planck-Institut f\"{u}r Kernphysik, Saupfercheckweg 1, 69117 Heidelberg, Germany}

\begin{abstract}
Galactic cosmic rays are believed to be accelerated at supernova remnants.
However, whether supernova remnants can be Pevatrons is still very unclear.
In this work we argue that PeV cosmic rays can be accelerated during the early phase of a supernova blast wave expansion in dense red supergiant winds.
We solve in spherical geometry a system combining a diffusive-convection equation which treats cosmic-ray dynamics coupled to magnetohydrodynamics to follow gas dynamics.
The fast shock expanding in a dense ionized wind is able to trigger the fast non-resonant streaming instability over day timescales, and energizes cosmic-rays even under the effect of p-p losses.
We find that such environments make the blast wave a Pevatron, although the maximum energy depends on various parameters such as the injection rate and mass-loss rate of the winds.
Multi-PeV energies can be reached if the progenitor mass loss rates are of the order of $10^{-3}$ M$_{\sun}$ yr$^{-1}$.
It has been recently invoked that, prior to the explosion, hydrogen rich massive stars can produce enhanced mass loss rates.
These enhanced rates would then favor the production of a Pevatron phase in early times after the shock breakout.
\end{abstract}


\section{Introduction}
More than a century after their discovery by V.F. Hess we do not still have a definite answer to the question of astrophysical sources where Cosmic Rays (CRs hereafter) are accelerated. It is widely accepted that a preferential mechanism of CR acceleration is associated with the diffusive shock acceleration (DSA) process (see an extensive review by \citet{Drury83}). 

In this process CRs gain energy by scattering off turbulent magnetic fluctuations up- and downstream the shock front. While accelerated CRs can carry a fraction of the order of 10\% of the kinetic inflow energy and are then able to trigger themselves the magnetic field fluctuations they need to complete successive Fermi cycles around the shock front.
Different instabilities have been invoked to be at the origin of the turbulence among which, the acoustic instability and the firehose instability both associated with anisotropic CR pressure ahead the shock front, the streaming instability induced by the drift motion of CRs in the background interstellar medium (see a review by \citet{Marcowith16}).
The streaming instability is anticipated to be important in fast moving shock waves where the non-resonant modes (i.e. perturbations with scale length smaller than the CR Larmor radius in the un-amplified interstellar magnetic field) have been shown to grow rapidly \citep{B04}.
The so-called non-resonant hybrid (NRH) instability (or Bell instability) for a given kinetic pressure imparted into the CRs has a maximum growth rate which scales as $ \xi\,  \sqrt{\rho}\,v_{\rm  sh}^3/E_{\rm esc}$, where $\xi$, $v_{\rm sh}$, $\rho$ and $E_{\rm esc}$ are the fraction of the kinetic gas energy imparted into CRs, the shock speed, ambient gas mass density, and the CR energy escaping from the system respectively.
Numerical simulations have intensively investigated the evolution of the magnetic field generation by the NRH instability \citep[e.g.][]{CS14c, BCSS15, VCM18, VCM19, HC19}.
These simulations showed that after a linear phase of growth, the instability saturates at a level $\delta B/B \sim M$, where $M$ is the shock (sonic/Alfv\'enic) Mach number.
In the non-linear phase of the magnetic field growth, the free energy is transferred to larger $k$ (wave number) modes which can become resonant with lower energy CR, i.e., their wave number verify $k\,\bar{R}_{\rm L}\sim 1$, where $\bar{R}_{\rm L}$ is the CR Larmor radius taken in the amplified magnetic field.

The maximum CR energy is fixed by the condition that the most energetic particles are able to carry a sufficient areal charge at the edge of the CR precursor, ahead the shock front \citep{B13}.
Then it can be shown that the maximum CR energy $E_{\rm max}$ scales as $\xi\,\sqrt{\rho}\,v_{\rm sh}^3\,t_{\rm age}$.
This relation and the shock Mach number dependence of the saturation magnetic field discussed above both point towards fast high Mach number shocks moving in dense media in order to produce the highest CR energies to possibly reach the CR knee energy around a few PeV \citep{SB13}.
Hence, the strong shock wave triggered after the explosion of core-collapse supernov\ae~can be considered as a potential PeVatron.
This is the first of the \emph{two} main assumptions in this work.
We base our calculations on works proposed by \citet{MDRTG18} and \citet{T09} where fast (with speed up to 0.1-0.2 $c$), strong (Mach numbers exceeding a few hundred) collisionless shocks pervading the dense circumstellar medium (CSM) of supernova (SN) massive progenitor are considered as ideal places to initiate a fast growth of NRH modes. In these works assuming typical saturation magnetic field strength by the NRH instability, the authors showed that  maximum CR energies may reach PeV within a few days after the shock breakout. The combination of a high shock speed and a high ambient density is indeed fulfilled in the earliest stage of the blast wave propagation in the CSM medium. It is then critical that the DSA can operate as early as possible in the shock expansion history (see, \citet{GB15} for a discussion). This our second main assumption.

In order to further investigate this issue, we conduct combined magnetohydrodynamic (MHD) and kinetic simulations using the code developed by \citet{I19}. The code has been upgraded to account for spherical geometry in order to capture the acceleration at the highest energies at best, which makes it possible to simulate CR acceleration under realistic parameters with the magnetic field amplification by the NRH instability. We complete these simulations by a series of numerical runs using a particle-in-cell-magnetohydrodynamic (PIC-MHD) technique developed by \cite{VCM18} in order to explore the early stage of the particle acceleration process. \\
The paper is organized as follows: in Section 2, we provide the basic equations and numerical settings for simulations. The results of the simulations and their physical interpretation are shown in Section 3. In Section 4 we discuss the PIC-MHD runs results, compare our results with previous studies of high Alfv\'enic Mach number shocks and then evaluate the implications of our results for observations. Section 5 summarizes the paper.

\section{Basic Equations and Numerical Setups}
\subsection{Basic Equations}\label{BE}
We solve a hybrid system of the Bell MHD equations and a telegrapher-type diffusion convection equations\footnote{Eq.~(\ref{DC2}) can be transformed into a telegrapher-type differential equation, if we take time-derivative of eq.~(\ref{DC2}) and substitute eq.~(\ref{DC1}).} in the polar coordinate around $\theta\sim \pi/2$ \citep{B13, I19}.
The hybrid system can be break into 1) gas dynamics equations:
\begin{eqnarray}
&& \frac{\partial\,\rho}{\partial t}+\frac{1}{r^2}\frac{\partial}{\partial r}r^2(\rho\,v_r)=0,\label{eq1}\\
&& \frac{\partial}{\partial t}(\rho\,v_r)+\frac{1}{r^2}\frac{\partial}{\partial r}r^2(\rho\,v_r^{2}+p+\frac{B_\theta^2+B_\phi^2}{8\,\pi})=0,\label{eq2}\\
&& \frac{\partial}{\partial t}(\rho\,v_\theta)+\frac{1}{r^2}\frac{\partial}{\partial r}r^2(\rho\,v_r\,v_\theta-\frac{B_r\,B_\theta}{4\pi})=-\frac{1}{c}j^{({\rm ret})}_{r}\,B_\phi,\label{eq3}\\
&& \frac{\partial}{\partial t}(\rho\,v_\phi)+\frac{1}{r^2}\frac{\partial}{\partial r}r^2(\rho\,v_r\,v_\phi-\frac{B_r\,B_\phi}{4\pi})=\frac{1}{c}j^{({\rm ret})}_{r}\,B_\theta,\label{eq4}\\
&& \frac{\partial\,\epsilon}{\partial t}+\frac{1}{r^2}\frac{\partial}{\partial r}r^2\{v_r\,(\epsilon+p+\frac{B^2}{8\,\pi}) -B_r\frac{ \vec{B}\cdot\vec{v} }{4\,\pi}\}=-\frac{j^{({\rm ret})}_{r}}{c}(\vec{v}\times\vec{B})_{r}\,, \label{eq5} \\
&& \epsilon=\frac{p}{\gamma-1}+\frac{1}{2}\rho\,v^2+\frac{B^2}{8\pi},\label{eq6}
\end{eqnarray}
2) magnetic field evolution:
\begin{eqnarray}
&& \frac{\partial\,B_\theta}{\partial t}=\frac{1}{r^2}\frac{\partial}{\partial r}r^2(B_r\,v_\theta-B_\theta\,v_r),\label{eq7}\\
&& \frac{\partial\,B_\phi}{\partial t}=\frac{1}{r^2}\frac{\partial}{\partial r}r^2(B_r\,v_\phi-B_\phi\,v_r),\label{eq8}
\end{eqnarray}
3) CR kinetic equations:
\begin{eqnarray}
&& \frac{\partial F_0(r,p)}{\partial t}+\frac{1}{r^2}\frac{\partial}{\partial r}\{r^2\,v_r\,F_0(r,p)\}-\frac{1}{3}\frac{\partial\,v_r}{\partial r}\frac{\partial\,F_0(r,p)}{\partial \ln p}\nonumber\\
&&\qquad =-\frac{c}{3}\frac{1}{r}\frac{\partial}{\partial r}\{r\,F_1(r,p)\}+Q_{\rm inj}(r,p)\,p^{3}-L_{\rm pp}(r,p)\,p^{3},\label{DC1}\\
&& \frac{\partial F_1(r,p)}{\partial t}+\frac{1}{r^2}\frac{\partial}{\partial r}\{r^2\,v_r\,F_1(r,p)\} \\
&&\qquad =-\frac{c}{r}\frac{\partial}{\partial r}\{r\,F_0(r,p)\}-\frac{c^2}{3\,\kappa(p,{\vec B})}F_1(r,p)-L_{\rm pp}(r,p)\,p^{3},\nonumber \label{DC2}
\end{eqnarray}
where $j^{({\rm ret})}_r$ is the return current density induced by the cosmic ray streaming current, i.e., $j^{({\rm ret})}_r=-j^{(\rm cr)}_r$, $Q_{\rm inj}$ is an injection rate, $L_{\rm pp}$ is a momentum loss rate due here to inelastic p-p collisions, and $\kappa$ is the diffusion coefficient, which generally depends on the momentum of the cosmic rays $p$ and local magnetic field.
In the polar coordinate, basic eqs.~(\ref{eq2})-(\ref{eq4}) and eqs.~(\ref{eq7})-(\ref{DC2}) must have the curvature terms that are inversely proportional to the radial coordinate $r$,
but in this paper, we neglect them because we consider a situation where the shock radius is very much larger than the scale of spatial derivatives (typically the growth scale of the NRH instability).
We also omit the effect of CR pressure on fluid dynamics, because we select a CR injection rate that does not cause substantial shock structure modification (see, \S \ref{pcr} for discussion).

Eqs.~(\ref{DC1}) and (\ref{DC2}) constitute the diffusion convection equation\footnote{One can easily confirm that these two equations recover the usual diffusion convection equation derived by \citet{S75}, if we take the limit $c\rightarrow\infty$.}, where $f_0(r,p)=F_0(r,p)/p^3$ is the isotropic component of the cosmic ray distribution function and $f_1(r,p)=F_1(r,p)/p^3$ is the anisotropic component so that the distribution function is given by $f(r,\vec{p})=f_0(r,p)+(p_r/p)\,f_1(r,p)$ (see Bell et al.~2013 for higher order equations\footnote{
Eqs.~(\ref{DC1}) and (\ref{DC2}) are from eqs.~(11a) and (11b) of \citet{B13}, where we neglect quadrupole term $f_{i\,j}$ and also $f_y$ and $f_z$ terms.
Due to the omission of $f_y$ and $f_z$, the $y$ and $z$ components of the cosmic ray current ($j_y, j_z$) are always set to be null.
$j_y$ and $j_z$ are induced when cosmic rays, whose gyro-radius is smaller than the wave-length of the magnetic field disturbance, stream along the disturbed field.
The current carried by these cosmic rays with small gyro-radius does not contribute to the growth of the NRH instability.
This is the reason why we limit the integration range of particle momentum in the current calculation given by eq.~(\ref{jcr2}).
}).
These CR kinetic equations are based on a multi-moment expansion formulation of the Vlasov equation derived by Bell et al. (2013).
Thus, no unphysical effect is introduced even if we employ these unfamiliar basic equations.
Thanks to the hyperbolic nature of the eqs.~(\ref{DC1}) and (\ref{DC2}), it is relatively easy to perform simulations employing modern parallel supercomputers.

The total cosmic ray current density is given by
\begin{eqnarray}
j_r(r)&=&e\,\int_{p_{\rm L}}^{p_{\rm U}} \frac{c}{3}\frac{p}{\sqrt{p^2+m_{\rm p}^2 c^2}}\,f_1\,4\,\pi\,p^2\,dp\nonumber\\&=&e\,\int_{p_{\rm L}}^{p_{\rm U}}\frac{4\,\pi\,c}{3}\frac{p}{\sqrt{p^2+m_{\rm p}^2 c^2}}\,F_1\,d\ln p\nonumber\\
&\equiv& \int_{p_{\rm L}}^{p_{\rm U}} j_p\,d\ln p, \label{jcr1}
\end{eqnarray}
where $p_{\rm L}$ and $p_{\rm U}$ are, respectively, the lower and upper boundary momenta considered in the simulation, and we have assumed that the cosmic rays are composed of protons.
To accurately calculate the cosmic-ray current that contributes to the NRH instability, we use the following current density instead of eq.~(\ref{jcr1}):
\begin{equation}\label{jcr2}
j^{(\rm cr)}_r(r)=\int_{p_{\rm B}}^{p_{\rm U}}j_p\,d\ln p,
\end{equation}
where the lower bound of the integral $p_{\rm B}$ is determined by the condition $p_{\rm B}c/e\,B=l_{\rm B, min}$ with B=$\sqrt{B_r^2+B_\theta^2+B_\phi^2}$ (see, \citet{I19} for the physical reason).
Here $l_{\rm B, min}=c\,B_r/4\pi\,j^{(\rm cr)}_r$ is the minimum scale of the NRH instability, and simple algebra yields to $p_{\rm B}=e\,B_{r}\,B/(4\pi\,j^{(\rm cr)}_r)$.
When $p_{\rm B}$ is not found in the range between $p_{\rm L}$ and $p_{\rm U}$, we set $j^{(\rm cr)}_r=0$.

We employ the following diffusion coefficient \citep{S75, CS14c}
\begin{equation}\label{kappa}
\kappa(p,{\vec B})=\frac{4}{3\,\pi}\frac{\max(B_r^2,\delta B^2)}{\delta B^2}\frac{v_{\rm CR}\,p\,c}{e\,\max(|B_r|,\delta B)},
\end{equation}
where $\delta B^2=B_\theta^2+B_\phi^2$, and $v_{\rm CR}$ is the cosmic-ray velocity at momentum $p$.
When magnetic field fluctuations are smaller than the mean field, i.e. $\delta B \leq B_{r}$, eq.~(\ref{kappa}) gives the diffusion coefficient due to pitch angle scattering, while it becomes the Bohm limit coefficient under the amplified field strength $\delta B$ when $\delta B>B_{r}$. This type of diffusion coefficient is supported from Particle-in-Cell (PIC) simulation \citep{CS14c} and test particle transport calculation in a super-Alfv\'enic turbulence \citep{RII16}.

We apply the following injection rate which assumes that a fraction $\eta$ of the thermal gas particles are injected into the acceleration process at momentum $p_{\rm inj}$ \citep{BGV}:
\begin{equation}
Q_{\rm inj}(r,p)=\frac{\eta\,n_{\rm ini}(r)\,v_{\rm sh}}{4\,\pi\,p_{\rm inj}^2}\delta (p-p_{\rm inj})\,\delta (r-r_{\rm sh}),
\end{equation}
where $n_{\rm ini}(r)$ is the initial density field of a CSM. The injection momentum $p_{\rm inj}$ is determined so that $\eta=\int_{p_{\rm inj}}^{\infty}\exp{(p^2/2\,m_{\rm g}\,k_{\rm B}\,T_{\rm d})}\,dp/\int_0^{\infty}\exp{(p^2/2\,m_{\rm g}\,k_{\rm B}\,T_{\rm d})}\,dp$, $n_{\rm ini}(r)$ is the upstream CSM number density, $T_{\rm d}\simeq 3\,m_{\rm g}\,v_{\rm sh}^2/16\,k_{\rm B}$ is the temperature of shock-heated gas, and $m_{\rm g}$ is the mean gas particle mass.
Note that we solve gas dynamics assuming $m=1.27\,m_{\rm p}$, while again CRs are treated as pure protons.
This treatment results in a underestimation of net CR charge and CR current roughly by 10\%.
Since it is numerically quite expensive to treat the momentum space from $p_{\rm inj}\,(\sim 0.1\,\mbox{GeV c}^{-1}$; the corresponding particle energy is $\sim 5$ MeV), we assume that CRs injected at $p_{\rm inj}$ obey the standard DSA process and are accelerated to $p_{\rm L}$ following the distribution of $f_{0}(r_{\rm sh})\propto p^{-4}$.
We will discuss the reliability of this assumption in section \ref{sec:PICMHD}.
Under this assumption, the injection rate of particles at $p=p_{\rm L}$ can be written as
\begin{equation}\label{Qinj}
Q_{\rm inj, num}(r,p)=\frac{\eta\,n_{\rm ini}(r)\,v_{\rm sh}\,p_{\rm inj}}{4\pi\,p_{\rm L}^3}\delta (p-p_{\rm L})\,\delta (r-r_{\rm sh}).
\end{equation}
Note that, for numerical implementation, the delta functions in the above formal expression should be replaced by reciprocals of momentum resolution $\Delta p$ at $p=p_{\rm L}$ and spatial resolution $\Delta x$ at shock front.

In the dense CSM, CRs lose their energy through inelastic p-p collisions \citep{MDRTG18}.
The momentum loss rate can be given by
\begin{equation}
L_{\rm pp}(r,p)=n(r)\,c\,\{\sigma_{\rm pp}(p)\,f_{0,1}(p)-\sigma_{\rm pp}(p/\epsilon_{\rm pp})\,f_{0,1}(p/\epsilon_{\rm pp})\},
\end{equation}
where $\sigma_{\rm pp}$ is a cross section, which we use a formula given by \citet{KATV14}, and $\epsilon_{\rm pp}$ is an energy loss fraction by a single collision which approximately takes a constant value of 0.5 thanks to the Feynman scaling \citep{KKM05}.

\subsection{Initial Conditions}\label{inicnd}
We employ a red supergiant (RSG) CSM model following Marcowith et al.~(2018).
The initial density field is
\begin{eqnarray}
\rho_{\rm ini}(r)&=&5\times 10^{-15}\mbox{ g cm}^{-3}\,\left( \frac{\dot{M}}{10^{-5}\,\mbox{M}_{\sun}\,\mbox{yr}^{-1}} \right)\,\left( \frac{r}{10^{14}\,\mbox{cm}} \right)^{-2}\nonumber\\
&&\times \,\left( \frac{v_{\rm wind}}{10\,\mbox{km s}^{-1}} \right)^{-1},
\end{eqnarray}
where $\dot{M}$ is a mass loss rate of the RSG wind, and $v_{\rm wind}$ is a wind velocity.
For the initial magnetic field, we assume that the RSG wind is highly turbulent and turbulent dynamo amplifies a seed field to the energy level $\varpi$ times the wind kinetic energy: 
\begin{equation}
B_{\rm ini}^2/8\pi=\varpi\,\rho_{\rm ini}\,v_{\rm wind}^2/2\ .
\end{equation}
This yields
\begin{eqnarray}
|B_{\rm ini}(r)|&=&0.25\,\varpi^{1/2}\,\mbox{ Gauss}\,\left( \frac{\dot{M}}{10^{-5}\,\mbox{M}_{\sun}\,\mbox{yr}^{-1}} \right)^{1/2}\,\left( \frac{r}{10^{14}\,\mbox{cm}} \right)^{-1}\nonumber\\
&&\times \,\left( \frac{v_{\rm wind}}{10\,\mbox{km s}^{-1}} \right)^{1/2}.\label{Bini}
\end{eqnarray}
The dynamo efficiency $\varpi$ is highly unknown. RSG winds seem to be driven by radial stellar pulsations \citep{B88}, which naturally induces turbulent flows invoking a high dynamo efficiency. Numerical simulations of turbulent dynamo show that the magnetic energy can grow to the comparable level to the kinetic energy of turbulence \citep{CV00, CVBLR09}. Given these backgrounds, we set $\varpi=1$ as fiducial value and search a smaller value case of $\varpi=0.2$. The model parameters used in our simulation models are summarized in Table 1. Because of this modeling, the Alfv\'enic Mach number takes value:
\begin{equation}
M_{\rm A}=1000\,\varpi^{-1/2}\left( \frac{v_{\rm wind}}{10\,\mbox{km s}^{-1}} \right)^{-1}\left( \frac{v_{\rm sh}}{10^4\,\mbox{km s}^{-1}} \right).
\end{equation}

\begin{figure}[t]
\includegraphics[scale=0.65]{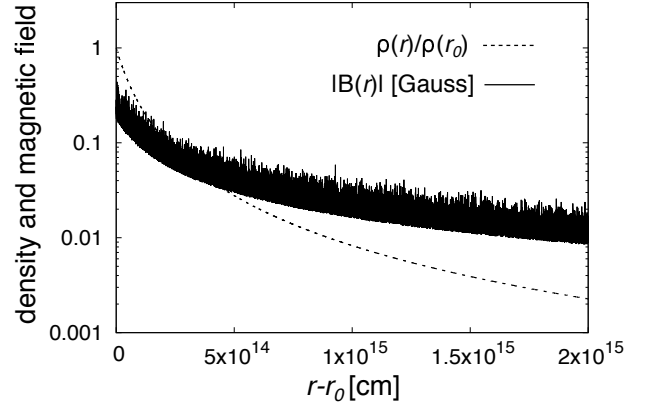}
\caption{\label{f1}
Initial CSM structure for Model 0.
Dashed line shows the density normalized at the inner boundary $r_0=10^{14}$ cm. 
Solid line shows the initial magnetic field strength $|B(r)|=\{B_r(r)^2+\delta B(r)^2\}^{1/2}$ for Model 0.
}
\end{figure}

The magnetic field estimated by eq.~(\ref{Bini}) is composed of a radial coherent component $B_{r}$ and fluctuation components $\vec{\delta B}=(B_{\theta},\,B_{\phi})$.
Since the turbulent dynamo often shows a flatter magnetic field power spectrum than that of the velocity field, we initially set a flat power spectrum of $\delta B_k^2 \propto k^0$ for simplicity.
The amplitude of $\vec{\delta B}$ is determined so that the energy is equally distributed in the radial and fluctuation components, i.e., 
$B_{r}^2(r)=\langle \delta B^2(r)\rangle =B_{\rm ini}^2(r)/2$.
In Figure \ref{f1}, we plot the structures of the initial density and magnetic field strength for the Model 0 (see table \ref{t1}).
To study the influence of the initial level of turbulent field, we also perform the quiet upstream case: $B_{r}^2(r)=B_{\rm ini}^2(r)/2,\,\langle \delta B^2(r)\rangle =0.01\times B_{r}^2(r)$.
Note that since we set radial dependence of $B_r$ to express the background field in our quasi-1D geometry, it infringes the divergence-free condition of the magnetic field.
In general, the divergence-free breakdown leads to an artificial Lorentz-force proportional to $\partial_r\,B_{r}(t=0)\sim B_r/r$.
However, influence of this penalty is very small, because, again, we treat the system with $1/r$ terms negligible compared to the radial derivative terms thanks to the very small-scale nature of the NRH instability.

Recent light curve study of type-II SNe showed that the mass-loss-rate of most RSG winds are enhanced by two orders of magnitude a few years to decades prior to the explosion producing CSM extending to $\gtrsim 10^{14}-10^{15}$ cm \citep{FMM18, OM19}. 
Thus, in this paper, in addition to the conventional CSM model made by a wind of $\dot{M}=10^{-5}\,\mbox{M}_{\sun}\,\mbox{yr}^{-1}$ (Model 0), we study denser CSM created by a wind of  $\dot{M}=10^{-3}\,\mbox{M}_{\sun}\,\mbox{yr}^{-1}$ suggested from the above observations (Model 1- Model 8, see table \ref{t1}).

Thermal pressure or temperature of the CSM is ambiguous. We take the initial thermal pressure so that the upstream sound speed becomes constant $c_{\rm s}(r)=100$ km s$^{-1}$, which leads constant sonic Mach number of $M_{\rm s}=100\,(v_{\rm sh}/10^4\mbox{ km s}^{-1})$.
We do not examine cases of different choice of the upstream thermal pressure, because physics of the NRH instability is not sensitive to the upstream thermal pressure
\footnote{One may have concerns about a possible effect of ion-neutral friction wave damping, which matters when the upstream temperature is not as high as fully ionizing the CSM.  However, according to eq.~(20) of \citet{Marcowith14}, the timescale of the friction damping in the present medium can be estimated to be much longer than the dynamical timescale of $\sim 10$ days.
Moreover, as discussed in \S 3.2.3 of \citet{MDRTG18} within the first weeks after the shock breakout the CSM has an ionization fraction close to one because the light elements are fully ionized by the blast wave X rays.}.

\subsection{Boundary Conditions}
To induce a blast wave shock, we set a cold ejecta at the inner spatial boundary at $r=r_0$ that takes $\rho(r_0)=10^3\,\rho_{\rm ini}(r_0)$ and $v_{r}(r_0)=v_{\rm eje}$.
In most models, we set $v_{\rm eje}=10,000$ km s$^{-1}$, which induces a blast wave shock of $v_{\rm sh}\simeq 14,000$ km s$^{-1}$.
For Model 5, to study a case of more energetic explosion, we set $v_{\rm eje}=20,000$ km s$^{-1}$ ($v_{\rm sh}\simeq 28,000$ km s$^{-1}$).

The inner spatial boundary radius is set at $r_0=10^{14}$ cm for Model 0, which is a few times larger than the RSG surface.
As for the high $\dot{M}$ models (Model 1 - Model 8), it is set at $r_0=10^{15}$ cm where $\rho_{\rm ini}(r_0)=5\times 10^{15}$ g cm$^{-3}$.
This is because a high CSM density region ($\rho\gtrsim 5\times 10^{-15}$ g cm$^{-3}$) is unavailable for particle acceleration due to the inelastic p-p collision loss.
The outer spatial boundary is set at $r_1=r_0+2.0\times 10^{15}$ cm for all the models, which enable us to follow a shock propagation more than 10 days for most of the models.

For the spatial boundaries of CRs, we assume that CRs do not penetrate into the ejecta.
Such a treatment can be implemented by setting $\kappa=0$ in the ejecta.
At the outer spatial boundary $r=r_1$, outgoing free boundary conditions are imposed: $F_0(p,r_1+\Delta x)=F_0(p,r_1)$ and $F_1(p,r_1+\Delta x)={\rm max}[\,F_1(p,r_1),\,0\,]$.
For the momentum space, we impose null values outside the numerical domain of $[p_{\rm L},\,p_{\rm U}]$.

\begin{table*} \label{t1}
\caption{Model Parameters}
\scalebox{1.}[1.]{
\begin{tabular}{c|cccccccc|c}
\hline\
Model ID & $\dot{M}$ [M$_{\sun}$ yr$^{-1}$] & $\varpi$ & $\langle \delta B_{\rm ini}^2(r)\rangle /B_{r,{\rm ini}}^2(r)$ & $\eta$ & $v_{\rm eje}$ [km s$^{-1}$] & p-p cooling & Bell term$^a$ & coordinate$^b$ & $E_{\rm cut}$\\  \hline
0 & $10^{-5}$ & 1.0 & 1.0 & $6\times 10^{-4}$ & $1.0\times10^4$ & yes & yes & polar & $0.8\times10^{15}$ eV ($t=7$ day) \\
1 & $10^{-3}$ & 1.0 & 1.0  & $6\times 10^{-4}$ & $1.0\times10^4$ & yes & yes & polar & $2.6\times10^{15}$ eV ($t=14$ day) \\
2 & $10^{-3}$ & 1.0 & 0.01 & $6\times 10^{-4}$ & $1.0\times10^4$ & yes & yes & polar & $2.3\times10^{15}$ eV ($t=14$ day) \\
3 & $10^{-3}$ & 1.0 & 1.0  & $6\times 10^{-4}$ & $1.0\times10^4$ & no & yes & polar & $5.0\times10^{15}$ eV ($t=14$ day) \\
4 & $10^{-3}$ & 0.2 & 1.0  & $6\times 10^{-4}$ & $1.0\times10^4$ & yes & yes & polar & $1.0\times10^{15}$ eV ($t=14$ day) \\
5 & $10^{-3}$ & 1.0 & 1.0  & $2\times 10^{-4}$ & $1.0\times10^4$ & yes & yes & polar & $1.3\times10^{15}$ eV ($t=14$ day) \\
6 & $10^{-3}$ & 1.0 & 1.0  & $6\times 10^{-4}$ & $2.0\times10^4$ & yes & yes & polar & $4.9\times10^{15}$ eV ($t=7$ day) \\
7 & $10^{-3}$ & 1.0 & 1.0  & $6\times 10^{-4}$ & $1.0\times10^4$ & no & yes & plane parallel & $7.0\times10^{15}$ eV ($t=14$ day)\\
8 & $10^{-3}$ & 1.0 & 1.0  & $6\times 10^{-4}$ & $1.0\times10^4$ & yes & no & polar & $1.6\times10^{14}$ eV ($t=14$ day) \\
\hline\
\end{tabular}
\tablenotetext{a}{if no, we always set null cosmic-ray current $j_{r}^{({\rm CR})}=0$.}
\tablenotetext{b}{In the plane parallel case, we set spatially constant upstream, whose physical values are same to those of $r=r_0$ cm in Model 1, and we solve basic equations in the plane parallel geometry.}
}
\end{table*}

\subsection{Domain Size and Resolution}\label{RES}
To capture the growth of the NRH instability, we need to at least resolve the most unstable scale of the instability \citet{B04}: $\lambda_{\rm Bell}=c\,B_{r}/|j_r^{({\rm CR})}|$.
As we will show detailed spatial distributions in the next section, the strength of the current density of Model 1, which we think the most realistic model, takes value of $j^{(\rm{cr})}_r\sim 0.01$ esu s$^{-1}$ cm$^{-2}$.
Thus, the most unstable spatial and timescales of the NRH instability can be estimated as (Bell 2004)
\begin{eqnarray}
\lambda_{\rm B}&\simeq& 3\times 10^{11} \mbox{ cm} \,\left( \frac{j^{({\rm cr})}}{0.01\,\mbox{esu s$^{-1}$cm$^{-2}$}} \right)^{-1}\left( \frac{B_r}{0.1\,\mbox{G}} \right), \label{kB}\\
\omega_{\rm B}^{-1}&=&\frac{\lambda_{\rm B}}{2\pi\,\,v_{\rm A}}=\frac{c\,\rho^{1/2}}{\pi^{1/2}\,|j^{(\rm{cr})}_r|} \nonumber\\
&\simeq& 0.6 \mbox{ day} \,\left( \frac{j^{({\rm cr})}}{0.01\,\mbox{esu s$^{-1}$cm$^{-2}$}} \right)^{-1}\left( \frac{\rho}{10^{-15}\,\mbox{g cm}^{-3}} \right)^{1/2}. \label{tB}
\end{eqnarray}
In order to resolve this scale with more than 10 numerical cells, we need a numerical cell number $N_{{\rm cell}, r}\gtrsim L_{\rm r}/(0.1\,\lambda_{\rm B})\sim 10^5$ indicating that a very high spatial resolution is required, where $L_{\rm r}=2\times 10^{15}$ cm is the spatial domain size.
To satisfy this requirement, we use a resolution $N_{{\rm cell},r}=2^{21}=2097152$ ($\Delta x=L/N_{{\rm cell},r}=0.95\times 10^{9}$ cm) for all the models.
For the momentum space, we consider the range $p_{\rm L}\,c=10^{12}\mbox{ eV}$ and $p_{\rm U}\,c=10^{16}\mbox{ eV}$, which is divided into uniform $N_{{\rm cell},p}=64$ numerical cells in the logarithmic scale, i.e., $\Delta \ln p=\ln (p_{\rm U}/p_{\rm L})/N_{{\rm cell},p}$.
As we show latter in Fig \ref{f32b}, most CR current is composed of CRs with $E>100$ TeV.
Thus, so far as $p_{\rm L}$ is selected below 100 TeV/$c$, choice of $p_{\rm L}$ would not likely affect the results.

\subsection{Numerical Scheme and Advantage of Telegrapher-type Basic Equations}
We employ the same numerical scheme that was developed by \citet{I19}.
Here, we briefly review its main structure:
The MHD equations are solved using the Godunov-type scheme with an approximate Riemann solver \citep{S99}.
The telegrapher-type diffusion convection equations are integrated using the fourth-order MUSCL scheme \citep{YD93}.
The source terms (RHSs of eq.~[\ref{eq3}] and eq.~[\ref{eq4}], the $L_{\rm pp}$ terms, and the damping term of $F_{1}$) are solved by using the piecewise exact solution method \citep{II08}.

A timestep of integration is determined by $\Delta t=C_{\rm CFL}\,\Delta r/(c/\sqrt{3})$, where $c/\sqrt{3}$ is the free streaming velocity of CRs.
The CFL number $C_{\rm CFL}$ is set to be 0.8.
Note that we do not need other timestep limiters, because the characteristic velocity originated in the MHD part hardly exceeds $c/\sqrt{3}$, and the piecewise exact solution method does not impose a time-step limitation.

As pointed out in \citet{I19}, there are big advantages to employ the telegrapher-type basic equations. One is the length of the timestep.
If we tackle the same problem by solving the conventional, parabolic, diffusion convection equation, the timestep imposed by an explicit scheme is $\Delta t_{\rm ex}=0.5\,\Delta x^2/\kappa(B,p)\sim 10^{-12}$ day for typical problem set ($B\sim 0.1$ Gauss and $p=p_{\rm U}$) that makes it impossible to integrate even over a day timescale.
On the other hand, our scheme impose $\Delta t=0.8\,\Delta r/(c/\sqrt{3})=5\times 10^{-7}$ day that is manageable with even more than 10 day timescale integration.

One may consider to employ an implicit scheme or super-time-stepping scheme to abbreviate the timestep issue.
However, the former one is incompatible with massive parallel computation, and the later one cannot provide orders of magnitude timestep extension in the present problem.

The other advantage is that the telegrapher-type equations appropriately handle both a diffusion regime of CR transport and a free streaming regime.
This nature is particularly important for capturing the CR current, because in a far upstream region, where the NRH instability grows, the current is mostly composed of nearly escaping high-energy CRs that are not confined by magnetic fields.

\subsection{Numerical Convergence}\label{cnv}

\begin{figure}[t]
\includegraphics[scale=0.6]{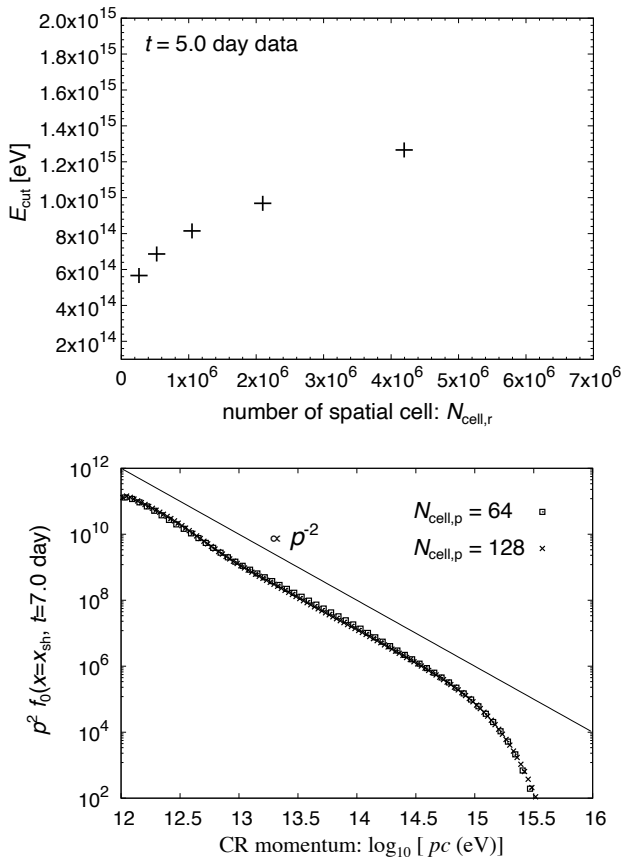}
\caption{\label{f2}
Top: cutoff energies at $t=5.0$ day as a function of spatial resolution.
Bottom: CR spectra at the shock front for Model 1 ($N_{{\rm cell},p}=64$) and the higher resolution run ($N_{{\rm cell},p}=128$).
To calculate spectra, the CR distribution function $f_0$ is spatially averaged from $r=r_{\rm sh}$ to $r_{\rm sh}-50\,\Delta r$.
}\end{figure}

As we will show in \S\ref{Sec:Rslt}, the unstable spatial scales of the NRH instability vary a lot with distance from the shock front.
The numerical resolution determined in \S\ref{RES} can be insufficient in particular at early stages and in the vicinity of the shock front.
Thus, we perform several simulations same to Model 1 with different resolution of $N_{{\rm cell}, r}=2^{18},\,2^{19},\,2^{20},\,2^{21},$ and $2^{22}$.
We also perform a larger momentum space resolution run with $N_{{\rm cell},p}=128$.

Since our main interest is the maximum energy achieved by the DSA, we assess the convergence through a cutoff energy at a fixed time.
The top panel of Figure~\ref{f2} shows the cutoff energies at $t=5.0$ day as a function of spatial resolution. Details of the dynamics will be given in the next section.
The cutoff energy is obtained by fitting CR spectrum $\log(f_0\,p^4)$ at the shock front through the least square method with a trial function $\log[A\,\exp\{-(p\,c/E_{\rm cut})^2\}]$\footnote{In the fitting, we use only the data larger than $0.1\,A$ for $f_0\,p^4$ to neglect the data having very small values.}.
Because a run with $N_{{\rm cell},x}=2^{22}$ was too expensive to continue more than $t=5.0$ day, we compare results at this time.
The plot shows that $E_{\rm cut}$ would have not yet reached convergence even at $N_{{\rm cell},r}=2^{22}$.
This would be due to unresolved magnetic field amplification at early stage where $B_{\rm ini}$ is larger than that used in eq.~(\ref{kB}) and near the shock front where $j^{({\rm CR})}$ is larger.

However, the curve seems to be becoming flat, and we can say that the converged $E_{\rm cut}$ would not differ more than factor two comparing to the resolution at $N_{{\rm cell},r}=2^{21}$.
Therefore, we claim that the results of Model 0 - Model 8 show lower values of the converged cutoff energy but error would be within a factor two.

As for the momentum space resolution, the results show perfect convergence thanks to a simple functional form of the resulting CR spectrum.
The bottom panel of Figure~\ref{f2} shows CR spectra at the shock front for Model 1 ($N_{{\rm cell},p}=64$) and the higher resolution run ($N_{{\rm cell},p}=128$) that are almost identical.

The interested reader can also refer to the results of more basic tests for the standard DSA and growth of the NRH instability in Appendix.

\section{Results} \label{Sec:Rslt}
\subsection{Basic Results of Model 0 and Model 1}
\subsubsection{Model 0: standard mass-moss rate}\label{sec:model0}

\begin{figure}[h]
\includegraphics[scale=0.85]{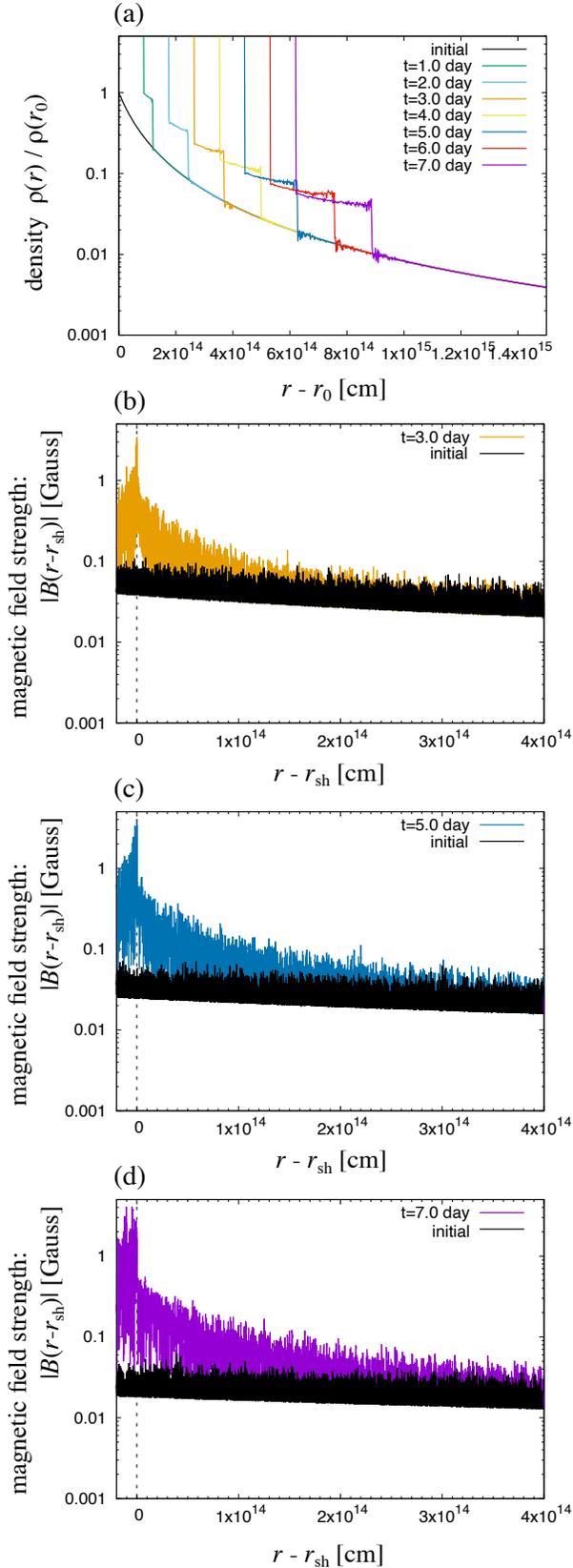}
\caption{\label{f31}
Panel (a): density structure normalized by $\rho(r_0)$ for Model 0.
Different colors show different snapshot times.
Panel (b): magnetic field strength around the shock front at $t=3.0$ day (orange). The initial structure is plotted as a purple line for a reference.
Dotted line shows shock position.
Panel (c): same as Panel (b) but for $t=5.0$ day.
Panel (d): same as Panel (b) but for $t=7.0$ day.
}\end{figure}

\begin{figure}[h]
\includegraphics[scale=0.7]{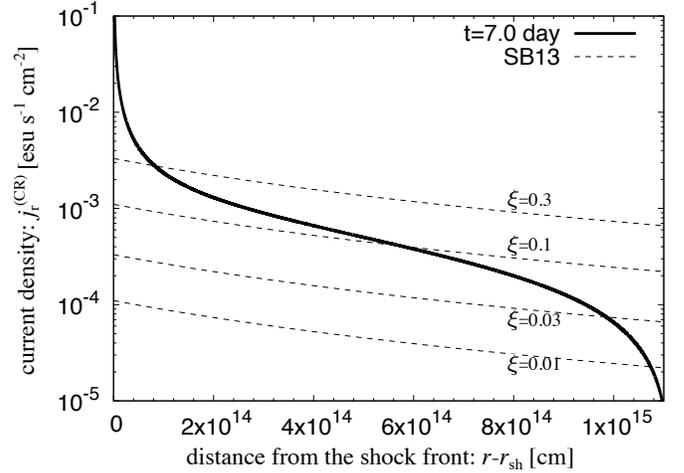}
\caption{\label{f32}
CR current density in the upstream region of the shock at $t=7.0$ day for Model 0.
Dashed lines show model current structures based on eq.~(3) of \citet{SB13}, where $\xi$ is the fraction of the kinetic gas energy imparted into CRs.
The model lines assume that only the geometrical effect attenuates the CR current density as $j_r\propto r^{-2}$.
}\end{figure}

\begin{figure}[h]
\includegraphics[scale=0.55]{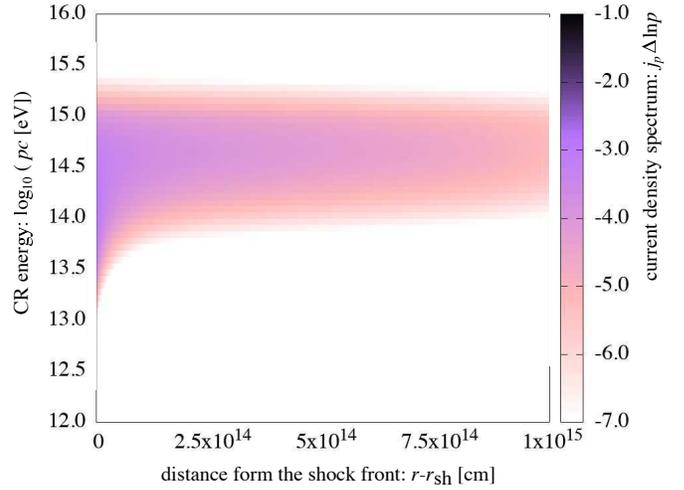}
\caption{\label{f32b}
CR current energy spectrum: $j_p\,\Delta \ln p$ (see, eq.~[\ref{jcr1}]) at $t=7.0$ day for Model 0.
}\end{figure}

Panel (a) of Figure \ref{f31} shows evolution of the density structure where we can read positions of the shock front and contact discontinuity, which separates the ejecta \& shocked CSM.
In Panels (b)-(d), we plot the structure of the magnetic field strength around the shock front.
The positions of the shock front at each time are evaluated as the maximum $r$ where $v_r>8\times 10^3$ km s$^{-1}$.
The black lines in these panels are the initial strength $|B(t=0,r)|$.
We see that the upstream magnetic field is amplified by the NRH instability by a factor of around ten at $r= r_{\rm sh}$.
Small scale fluctuations are observed even in the upstream density structure, although it can be seen only at very close to the shock front.
This can be explained by back-reaction of Alfv\'en waves induced by the NRH instability, because Alfv\'en velocity evaluated by using amplified field strength at near the shock front is supersonic and as large as 10\% to the forward shock velocity.

To see detail of the NRH instability, we plot the CR current density in the upstream region of the shock at $t=7.0$ day in Figure \ref{f32}.
We also plot the CR current energy spectrum $j_p\,\Delta \ln p$ in Figure \ref{f32b} at $t=7.0$ day, showing that high-energy escaping CRs with $pc\sim 10^{14}-10^{15}$ eV constitute the CR current. 
Comparing the structures of the magnetic field and the current density in Figure \ref{f32}, we find that the NRH instability is effective in a region with $j_{r}^{({\rm CR})}\gtrsim 10^{-3}$ esu s$^{-1}$ cm$^{-2}$.
In many previous theoretical modelings, it have been assumed that the NRH instability amplifies magnetic field until it reaches the so called saturation level\footnote{The saturation happens once the gyro radius of maximum energy CRs becomes smaller than the NRH instability critical scale.} \citep{B04}:
\begin{eqnarray}
B_{\rm sat}&=&\left(\frac{4\pi\,j_{r}^{({\rm CR})}\,p_{\rm max}}{e}\right)^{1/2}\nonumber\\
&\simeq& 1.2\,\mbox{Gauss}\,\left(\frac{j_{r}^{({\rm CR})}}{10^{-3}\,\mbox{esu s}^{-1}\,\mbox{cm}^{-2}}\right)^{1/2}
\,\left(\frac{p_{\rm max}}{1\,\mbox{PeV}\,c^{-1}}\right)^{1/2}, \label{eq:Bsat}
\end{eqnarray}
where $p_{\rm max}$ is the maximum (escaping) CR momentum.
It is clear that the level of the magnetic field in the simulation is smaller than this saturation level.
Note that a CR diffusion length is given by 
\begin{eqnarray}
l_{\rm diff}&=&\kappa(p)/v_{\rm sh}\nonumber\\
&\simeq& 10^{14}\mbox{ cm}\,\,\xi_B^{-1}\left(\frac{p_{\rm max}}{1\,\mbox{PeV}\,c^{-1}}\right) \left(\frac{B}{0.3\,\mbox{G}}\right)^{-1} \left(\frac{v_{\rm sh}}{10^4\,\mbox{km s}^{-1}}\right)^{-1},
\end{eqnarray}
where $\xi_B=\delta B^2/B^2$.
In order for the magnetic field to confine high energy CRs, the magnetic field should be amplified in the upstream region over $l_{\rm diff}$ from the shock front.

In Figure \ref{f32}, we plot the current structure based on eq.~(3) of \citet{SB13} as dashed lines, where the parameter $\xi$ is the fraction of the kinetic gas energy imparted into CRs.
In their model, it is assumed that only the geometrical effect attenuates the CR current density as $j_r\propto (r/r_{\rm sh})^{-2}$.
However, the result of our simulation shows that it drops more rapidly in particular at $r-r_{\rm sh}\gtrsim 10^{15}$ cm.
This stems from the fact that we cannot fill all upstream region by the escaping CRs, since the CR escape starts from a finite past, emphasizing the importance for solving the temporal evolution of the CR current.

\begin{figure}[h]
\includegraphics[scale=0.7]{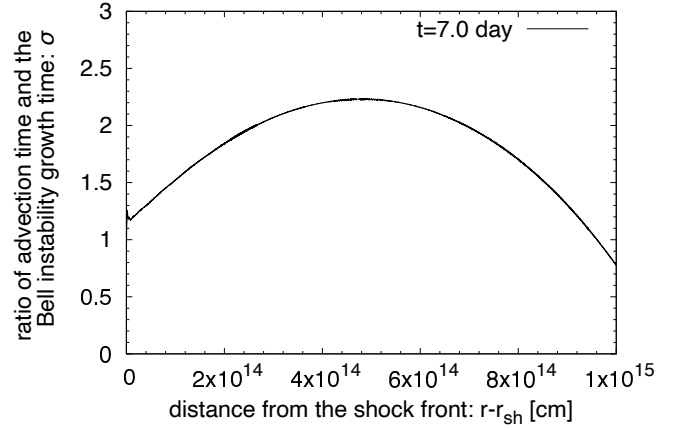}
\caption{\label{f33}
Ratio of advection time ($\{r-r_{\rm sh}\}/v_{\rm sh}$) and the NRH instability growth time ($\omega_{\rm B}^{-1}$) as a function of distance form the shock front at $t=7.0$ day.
}\end{figure}

The finite radial extent of the CR current is directly connected with the non-saturation of the magnetic field amplification.
To show this, we plot the ratio of an advection time ($\{r-r_{\rm sh}\}/v_{\rm sh}$) and the NRH instability growth time ($\omega_{\rm B}^{-1}$) as a function of distance form the shock front in Figure \ref{f33}.
The ratio is expressed as
\begin{eqnarray}
\sigma=\omega_{\rm B}\,t_{\rm adv}=\frac{\pi^{1/2}\,|j_r^{(\rm{CR})}|\,(r-r_{\rm sh})}{c\,\rho^{1/2}\,v_{\rm sh}}.\label{eqsig}
\end{eqnarray}
Figure \ref{f33} clearly shows that the ratio has a peak at $\sigma\simeq 2.2$, indicating that the upstream magnetic fields only have about two growth times until it is advected to the shock front, which is not enough to reach the saturation level.
Note that we can regard the ratio $\sigma$ as an e-folding number of the NRH instability, and indeed $\exp(\sigma=2.2)\simeq 9$ gives a good estimate of the amplification level.

\begin{figure}[h]
\includegraphics[scale=0.7]{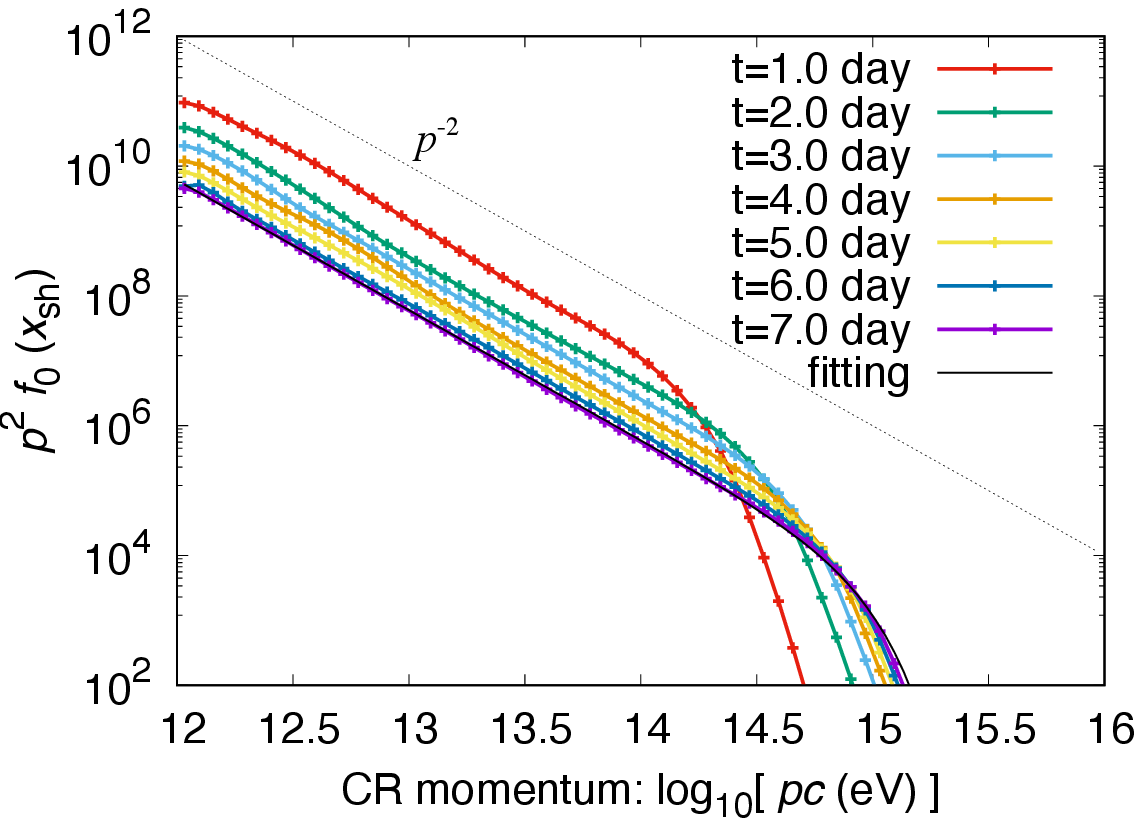}
\caption{\label{f34}
CR spectra of Model 0 around the shock front.
To calculate spectra, the CR distribution function $f_0$ is spatially averaged from $r=r_{\rm sh}$ to $r_{\rm sh}-50\,\Delta r$.
The resulting spectra at $t=7.0$ day is well fitted by $f_0\propto p^{-4}\,\exp\{-(p\,c/E_{\rm cut})^2\}$ (see \S \ref{cnv} for the fitting methodology) with $E_{\rm cut}=0.79\times10^{15}$ eV, which is plotted as black line.
}\end{figure}

According to the standard DSA, by imposing an acceleration timescale equaled to the shock age, we can estimate the maximum energy of CRs as
\begin{equation}
E_{\rm max}\simeq 0.4\times 10^{14}\mbox{ eV}\,\xi_B \left(\frac{B}{0.03\,\mbox{G}}\right) \left(\frac{v_{\rm sh}}{10^4\,\mbox{km s}^{-1}}\right)^2 \left(\frac{t}{10\,\mbox{day}}\right).
\end{equation}
In the above estimate, we substitute a non-amplified magnetic field level that leads the maximum energy far below 1 PeV.
The result of the simulation shows larger $E_{\rm max}$ thanks to the NRH instability.
Figure \ref{f34} is the resulting CR spectra of Model 0 around the shock front.
The fitting of the spectrum at $t=7.0$ day shows that $E_{\rm cut}=0.79\times 10^{15}$ eV.
Given that the higher spatial resolution leads roughly factor two larger $E_{\rm cut}$ (see \S \ref{cnv}), we can claim that the SNR similar to Model 0 would be a PeVatron.

In Model 0, we have terminated the run at $t=7.0$ day, at which the shock is propagating at $r-r_0\simeq 0.9\times 10^{15}$ cm.
Even if we continue the simulation, we hardly get higher $E_{\rm cut}$, because the background upstream magnetic field level drops down to $\sim 0.01$ Gauss.

\subsubsection{Model 1: higher mass loss rate}\label{sec:model1}

\begin{figure}[h]
\includegraphics[scale=0.85]{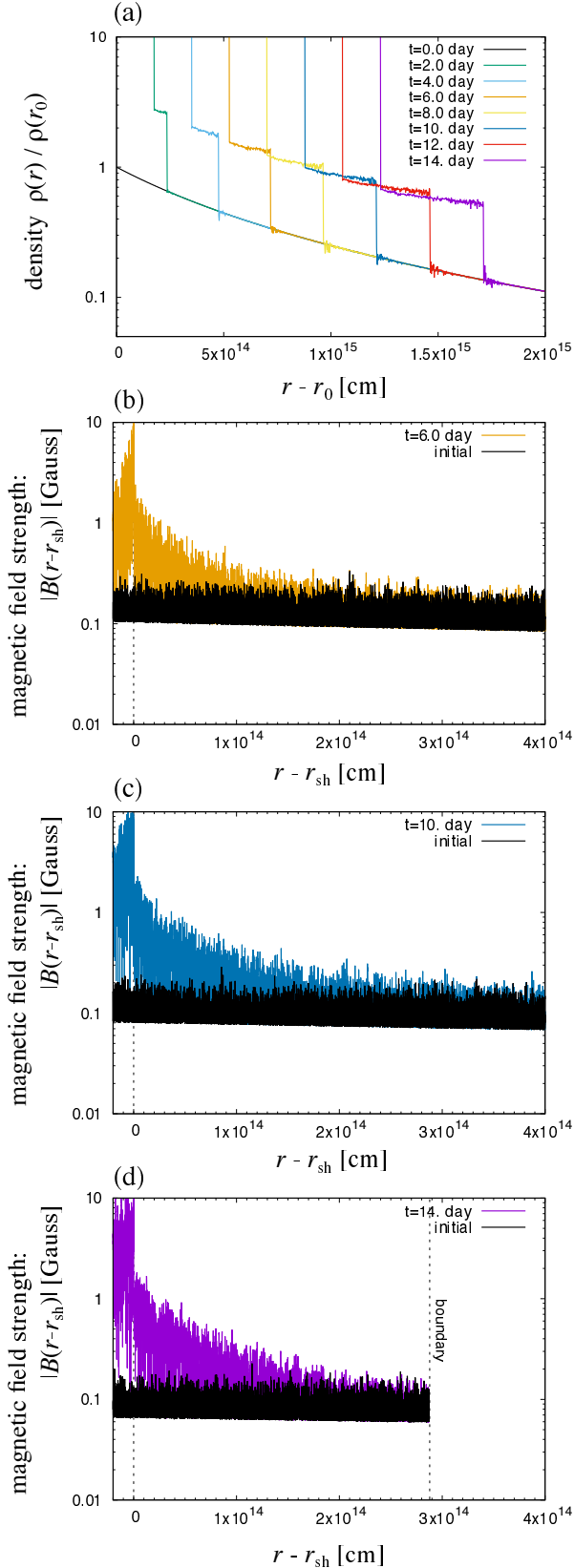}
\caption{\label{f41}
Panel (a): density structure normalized by $\rho(r_0)$ for Model 1.
Different colors show different snapshot times.
Panel (b): magnetic field strength around the shock front at $t=6.0$ day (orange). The initial structure is plotted as a purple line for a reference.
Dotted line shows shock position.
Panel (c): same as Panel (b) but for $t=10.0$ day.
Panel (d): same as Panel (b) but for $t=14.0$ day.
}\end{figure}

\begin{figure}[h]
\includegraphics[scale=0.7]{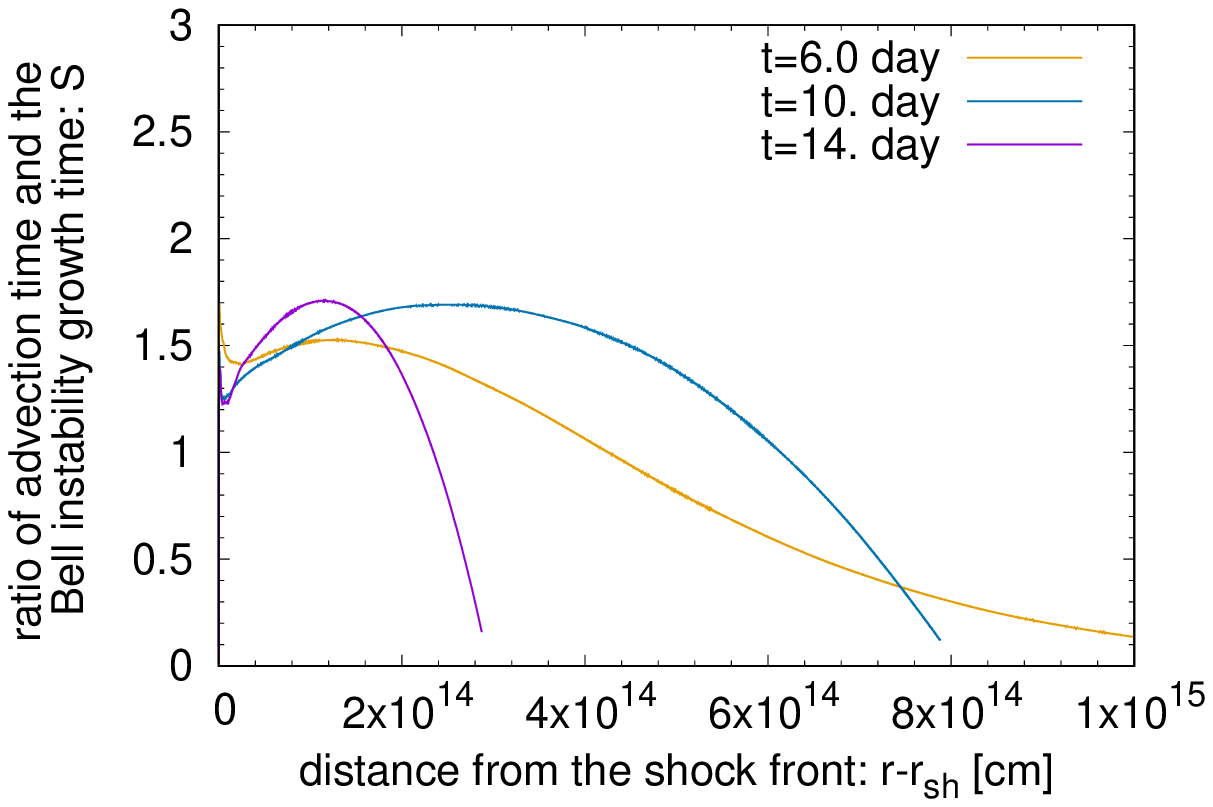}
\caption{\label{f43}
Ratio of advection time ($\{r-r_{\rm sh}\}/v_{\rm sh}$) and the NRH instability growth time ($\omega_{\rm B}^{-1}$) as a function of distance form the shock front at $t=10.0$ day for Model 1.
}\end{figure}

\begin{figure}[h]
\includegraphics[scale=0.7]{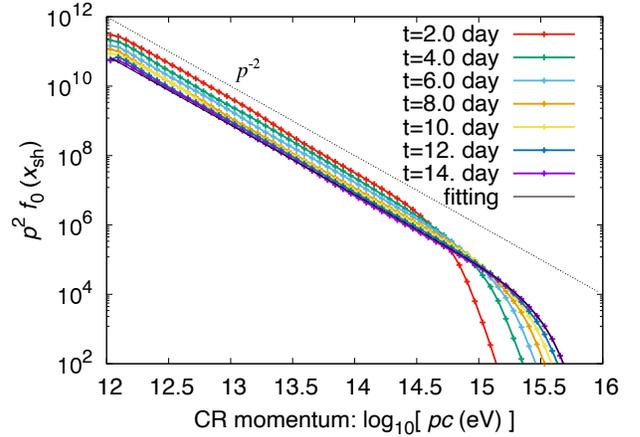}
\caption{\label{f44}
CR spectra of Model 1 around the shock front.
To calculate spectra, the CR distribution function $f_0$ is spatially averaged from $r=r_{\rm sh}$ to $r_{\rm sh}-50\,\Delta r$.
The resulting spectra at $t=14.0$ day is well fitted by $f_0\propto p^{-4}\,\exp\{-(p\,c/E_{\rm cut})^2\}$ (see \S \ref{cnv} for the fitting methodology) with $E_{\rm cut}=2.6\times10^{15}$ eV, which is plotted as black line.
}\end{figure}

As mentioned in \S \ref{inicnd}, recent observations show that a high mass-loss-rate wind model with $\dot{M}\sim 10^{-3}$ M$_{\sun}$ yr$^{-1}$ is plausible as a RSG CSM in particular $r\lesssim$ a few $\times 10^{15}$ cm \citep{FMM18}.
We study such high mass-loss-rate CSM models in Model 1-9, and here we discuss the result of Model 1 as a fiducial model.

Figure \ref{f41} shows the similar plots as Figure \ref{f31} but data is from Model 1.
We also plot the ratio $\sigma$ for Model 1 at $t=6.0,\,10.0,\,$ and 14.0 day in Figure \ref{f43}.
The result of Model 1 shows larger upstream magnetic field than that of Model 0.
This is due to the difference of the initial CSM condition, because the influence of the NRH instability is similar to that of Model 0, i.e., $\sigma$ takes similar value to that of Model 0 (see, Figure \ref{f43}).
In this model, we stopped simulation at $t=14.0$ day that is longer than that of Model 0, because the initial background magnetic field level is higher than Model 0 leading to a longer time evolution of $E_{\rm cut}$.
Figure \ref{f44} shows the CR spectra of Model 1 around the shock front, exhibiting more energetic particles than Model 0.
The fitting of the spectrum at $t=14.0$ day shows that $E_{\rm cut}=2.6\times 10^{15}$ eV, almost reaching the knee energy.
Even at $t=7.0$ day, $E_{\rm cut}=1.3\times 10^{15}$ eV, roughly twice larger than that of Model 0
\footnote{At $t=14$ day, the blast wave shock is propagating at $r\simeq r_0+v_{\rm sh}\,t \simeq 2.7\times10^{15}$ cm, which can be larger than the spatial extent of the CSM created by the high $\dot{M}$ wind (a few times $10^{15}$ cm). If so, the maximum energy obtained in Model 1-8 would be limited by the spatial extent of the dense CSM.}.

To make it clear the effect of the NRH instability, we have performed a simulation without the NRH instability as Model 8, which is done by artificially setting $j_{r}^{({\rm CR})}=0$.
From a fitting of the resulting CR spectrum at $t=14.0$ day, we obtain $E_{\rm cut}=1.6\times 10^{14}$ eV.
Given that the upstream magnetic field is amplified by an order of magnitude in Model 1, a factor $\sim 10$ larger $E_{\rm cut}$ in Model 1 than that of Model 8 is reasonable.

\subsection{Model 2: Smaller Initial $\delta$B Case}

\begin{figure}[h]
\includegraphics[scale=0.85]{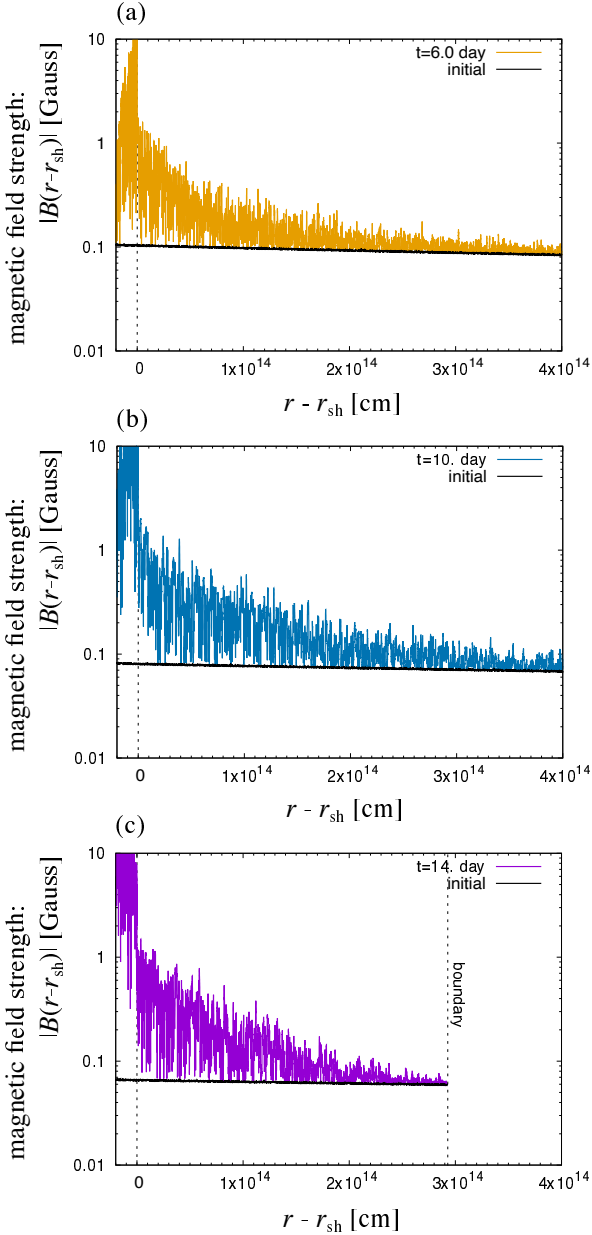}
\caption{\label{f51}
Panel (a): magnetic field strength around the shock front at $t=6.0$ day (orange) for Model 2. The initial structure is plotted as a purple line for a reference.
Dotted line shows shock position.
Panel (b): same as Panel (b) but for $t=10.0$ day.
Panel (c): same as Panel (b) but for $t=14.0$ day.
}\end{figure}

\begin{figure}[h]
\includegraphics[scale=0.7]{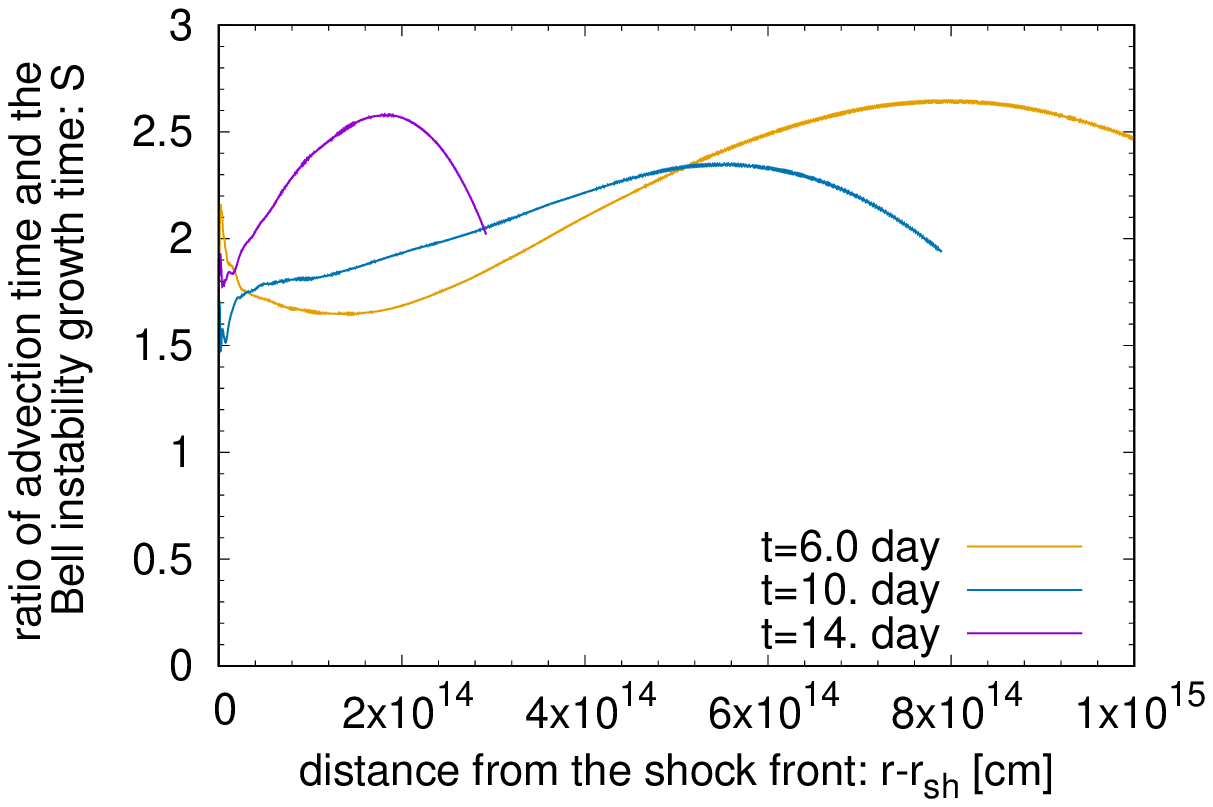}
\caption{\label{f53}
Ratio of advection time ($\{r-r_{\rm sh}\}/v_{\rm sh}$) and the NRH instability growth time ($\omega_{\rm B}^{-1}$) as a function of distance form the shock front at $t=10.0$ day for Model 2.
}\end{figure}

\begin{figure}[h]
\includegraphics[scale=0.7]{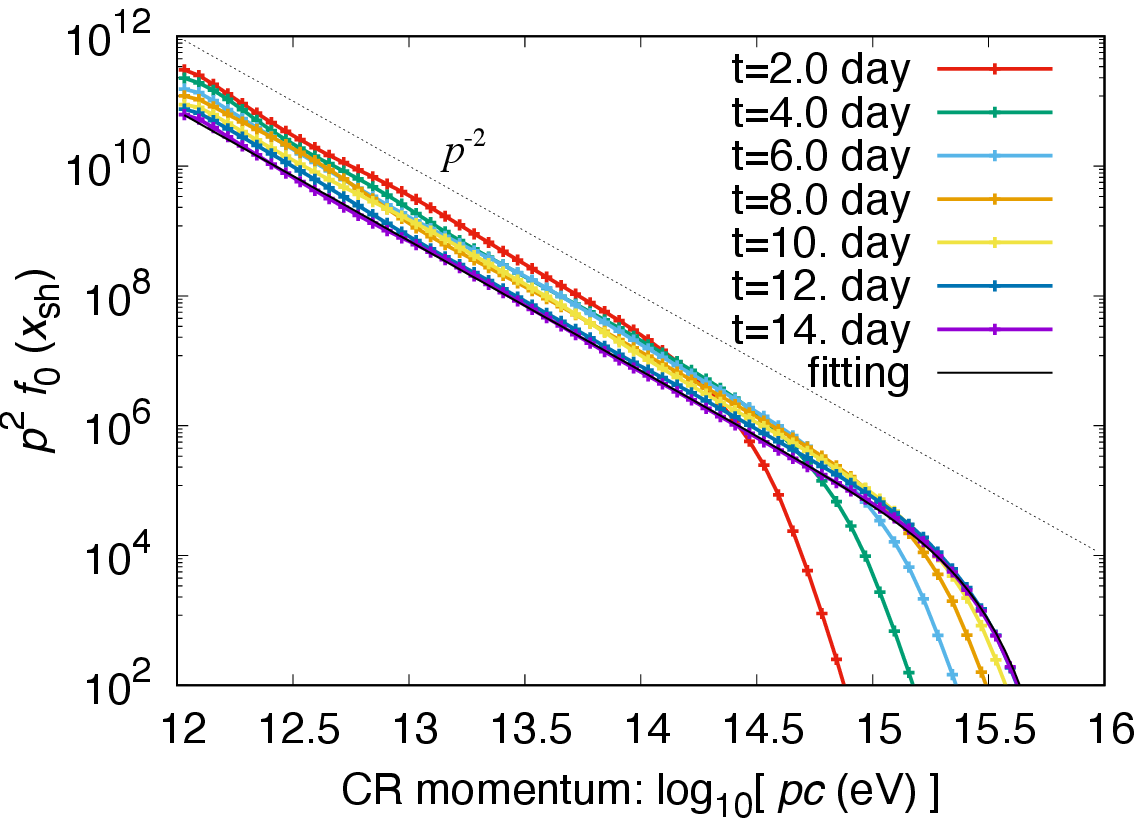}
\caption{\label{f54}
CR spectra of Model 2 around the shock front.
To calculate spectra, the CR distribution function $f_0$ is spatially averaged from $r=r_{\rm sh}$ to $r_{\rm sh}-50\,\Delta r$.
The resulting spectra at $t=14.0$ day is well fitted by $f_0\propto p^{-4}\,\exp\{-(p\,c/E_{\rm cut})^2\}$ (see \S \ref{cnv} for the fitting methodology) with $E_{\rm cut}=2.3\times10^{15}$ eV, which is plotted as black line.
}\end{figure}

In Model 2, we study a model with smaller initial $\delta B$, but the same $B_r$ as Model 1.
Since the original NRH instability assumed a coherent background magnetic field ($B_r\gg\delta B$), it is meaningful to study such a case.
Figure \ref{f51} is composed of the similar plots as panels (b)-(d) of Figure \ref{f31} but data is from Model 2.
The magnetic field is amplified almost the same level to Model 1 that can be understood from the ratio $\sigma$ plotted in Figure \ref{f53}.
The figure shows a larger value of $\sigma$ with a farther peak from the shock compared to Figure \ref{f43}, indicating more active growth of the magnetic field than Model 1 in particular at earlier stages.
From eq.~(\ref{eqsig}), this larger $\sigma$ is due simply to a larger $j_r^{({\rm CR})}$, and that is a consequence of the smaller initial $\delta B$, because the smaller $\delta B$ makes it easier for CRs to stream away.
So far as the NRH instability determines the label of $\delta B$, the final level of magnetic field strength does not substantially depends on the initial level of $\delta B$, which is reasonable as long as we use an injection model that is independent of the initial $\delta B$.
The resulting CR spectra are shown in Figure \ref{f54}.
The cutoff CR energy at $t=14.0$ day is obtained to be $E_{\rm cut}=2.3\times10^{15}$ eV that is similar to that of Model 1 as expected.

\subsection{Effect of p-p cooling}
One may wonder why that $E_{\rm cut}$ of Model 0 (0.8 PeV) and that of Model 1 (1.3 PeV) at $t=7.0$ day differ roughly only by a factor 2, even though the initial magnetic field strengths at $r-r_0=0.9\times 10^{14}$ cm ($\sim$ shock position at $t=7.0$ day) differ roughly factor 5.
The effect of the inelastic p-p collision cooling is filling the gap, because the density at $r-r_0=0.9\times 10^{14}$ cm in Model 1 is $\sim 28$ times larger than that in Model 0 owing to the high $\dot{M}$.
The resulting $E_{\rm cut}$ in Model 3 (no p-p cooling run) at $t=7.0$ day and 14.0 day are 2.5 PeV and 5.0 PeV, respectively.
Thus, we can conclude that the effect of the p-p collision cooling reduces $E_{\rm cut}$ by roughly factor 2 for the high $\dot{M}$ models.

Because of our choice of the boundary conditions, the number conservation of the CRs can be broken through the boundary leakage mostly at $p=p_{\rm L}$ and $x=L_{\rm box}$.
This leakage leads an underestimated CR current and growth rate of the NRH instability.
The result of Model 3 is ideal to check the number conservation.
We have counted the total number of injected particles and compared it with the total number of CRs in the numerical domain.
At $t=10$ day, the fraction of the leaked CRs through the boundaries is only 3.0\% to the total number of the numerically injected particles at $p=p_{\rm L}$.
At the final time of $t=14$ day, it rises to 8.5\%, because the shock is approaching the boundary.
Thus, even if all of these numerically leaked CRs were escaping CRs, the error on the CR current from the leakage is less than 10\%.

\subsection{Other Parameters Survey}
In Model 4, we reduced the parameter $\varpi$ to 0.2, which means the CSM magnetic energy is 20\% to the kinetic energy of the wind.
In Model 5, the injection rate is reduced to one-thirds of the other models, which dampens the cosmic-ray current and thus weakens the magnetic field amplification.
The resulting cutoff energies are summarized in the rightmost row of Table \ref{t1}.
We see smaller $E_{\rm cut}$ for Model 4 and Model 5 compared to Model 1 as expected.
Model 6 is the larger shock velocity case that simply enhances $E_{\rm max}$ at a fixed shock traveling distance, and we can confirm larger $E_{\rm cut}$ in Table \ref{t1}.

Finally we can study the influence of spherical geometry, if we compare results of Model 3 and Model 7.
In these two models, the effect of the p-p collision cooling is switched off, and only geometrical conditions are different.
In Model 7, we set a plane parallel geometry, i.e., the $r$ dependence of the initial condition is omitted and we solve the Cartesian coordinate version of the basic equations.
The result shows that the effect of curvature ($r$ dependence of the initial condition and diverging outward CR flux that dilutes escaping CR current density) reduces $E_{\rm cut}$ by $\simeq 30$\%.

\begin{figure}[h]
\includegraphics[width=\columnwidth]{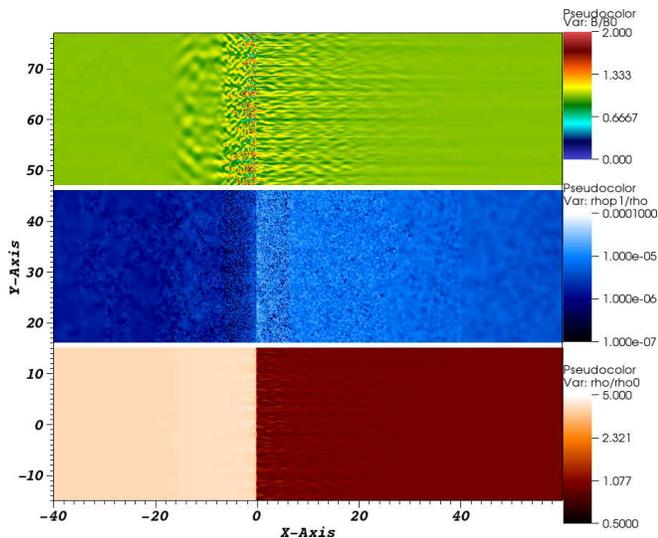}
\caption{\label{fig:picmhd1}
$M_{\rm a}=1000$ shock solutions at t = 180 $\omega_{\rm inj}$ using the PIC-MHD code: up turbulent magnetic field amplitude $B/B_0$, middle: CR density relative to the the thermal gas density, bottom: gas mass density relative to the initial upstream density. This early stage image shows the start of the filamentary structure in the upstream medium
}\end{figure}
\begin{figure}[h]
\includegraphics[width=\columnwidth]{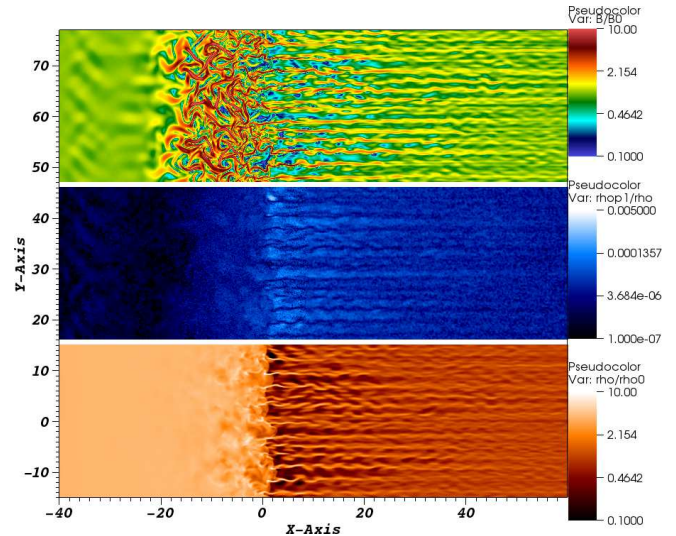}
\caption{\label{fig:picmhd2}
Identical to figure \ref{fig:picmhd1} but at t = 450 $\omega_{\rm inj}$. By this stage, the filaments have become more pronounced and start to influence the shape of the shock front.
}\end{figure}
\begin{figure}[h]
\includegraphics[width=\columnwidth]{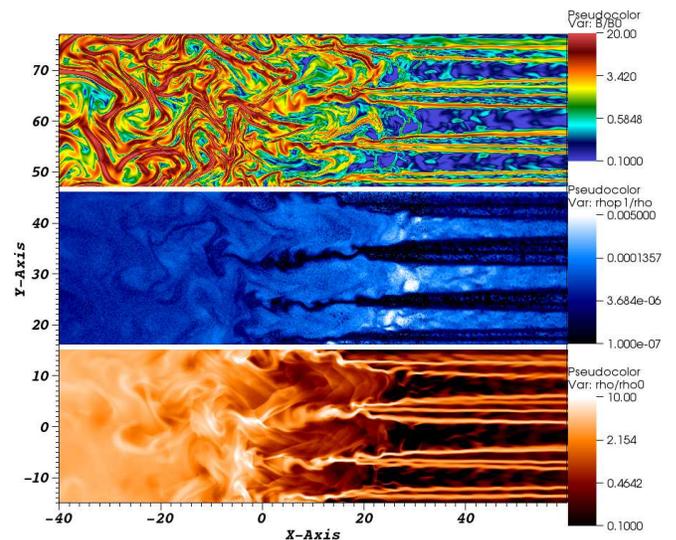}
\caption{\label{fig:picmhd3}
Identical to figure \ref{fig:picmhd1} but at t = 900 $\omega_{\rm inj}$. The difference in ram pressure at the shock front has caused large-scale corrugation of the shock.
}\end{figure}

\section{Discussion}

\subsection{PIC-MHD model}\label{sec:PICMHD}
One important assumption of the kinetic simulations performed in this work is that CR are injected at relativistic energies of 1 TeV. This is a strong assumption which relies on another assumption of efficient particle injection into the DSA process in these environments. In order to test this hypothesis we perform a numerical simulation of a high-Mach parallel shock using the PIC-MHD method \citep{BCSS15,VCM18}. This approach combines elements of both traditional MHD and PIC. The plasma is divided into two components: the first is the thermal plasma, which is treated as a fluid using grid-based MHD, the second is the non-thermal plasma, which is treated as a collection of particles that are modelled using the PIC method. The two fluids interact with each other through the electromagnetic field using a modified version of Ohm's law. The code is based on the {\tt{MPI-AMRVAC}} code \citep{VKM08}. This is a fully conservative finite volume code that solves the conservation equations of MHD on an adaptive mesh. 

We start our simulation from the Rankine-Hugoniot conditions of a standing shock with $M_A=1000$ and upstream velocity $v_{\rm s}=0.03\,c$. Once the simulation starts, we inject particles at the shock with an injection rate of $2\times10^{-3}$ of the total mass flowing through the shock.  This rate has been chosen to allow a growth of the NRH instability over reasonable computational timescales. The particles are given a starting velocity $v_{\rm inj}\,=\,3 v_s$, with a direction that is randomly chosen to create an isotropic distribution in the post-shock restframe. These values match the ones used previously in \citet{VCM18,VCM19} as well as the precition used by \citet{BCSS15}. While this is higher than the injection rate used in previous sections, it is unlikely to influence the results. Comparing equations B8 and B5 from \citet{V20} gives us the minimum upstream particle density required to induce the non-resonant streaming instability. For the simulation parameters used here, this results in a relative particle to gas density of $\rho_{p}/\rho_0\,=\,3.3\times10^{-5}$. Because the injection takes place anisotropically, this gives us a minimum injection rate $6.6\times10^{-5}$, which we exceed in every case.
For our simulation box, we use a 2-D grid that is $480\times 30 R_l$, with $R_l$ the gyro radius of the particles at injection. This space is covered by a grid that has $480\times30$ grid cells at its coarsest level. The code uses adaptive mesh refinement, allowing four addition levels for a maximum effective resolution of $0.0625\,R_l$ per grid cell.  

The results of our simulations are shown in Figs~\ref{fig:picmhd1}-\ref{fig:picmhd3}, which show (from top to bottom) the change in the magnetic field ($B/B_0$), the relative density of the non-thermal particles ($\rho_{nt}/\rho$) and the thermal gas density at $t=180$, 450, and $900\,,R_l/v_{\rm inj}$. 
 From an early stage, the current generated by the movement of the non-thermal particles in the upstream medium causes a disturbance in the upstream magnetic field (Fig.~\ref{fig:picmhd1}), creating filaments along the direction of the flow where the magnetic field strength is amplified.
This disturbance grows over time, and becomes visible in both the thermal and non-thermal gas density distribution as they respond to the changes in the magnetic field (Fig.~\ref{fig:picmhd2}). The thermal gas, which is assumed to be fully ionized, is coupled directly to the magnetic field, causing the thermal gas density to form filaments that coincide with the magnetic field amplification. However, the particles, which have a greater freedom of motion owing to their high velocity can travel more easily in areas where the magnetic field is weak and relatively smooth. As a result, the non-thermal particle density is highest in the low-density regions. (N.B. this effect is enhanced in Figs.~\ref{fig:picmhd1}-\ref{fig:picmhd3} because we plot the density of the non-thermal plasma relative to the thermal gas density rather than the absolute density)
Eventually, this disturbance of the upstream medium creates a variation in the ram pressure at the shock front, causing the shock to become corrugated (Fig.~\ref{fig:picmhd3}). This process was previously shown using the same code in \citet{VCM18}, which showed that these are the result of the  non-resonant streaming instability, as well as in various other models using both PIC-MHD and PIC-hybrid methods \citet[e.g.][]{CS14b,CS14c,BCSS15}. \citet{VCM19} confirmed that the same pattern occurs if the simulation is run in 3-D.

\begin{figure}
\includegraphics[width=\columnwidth]{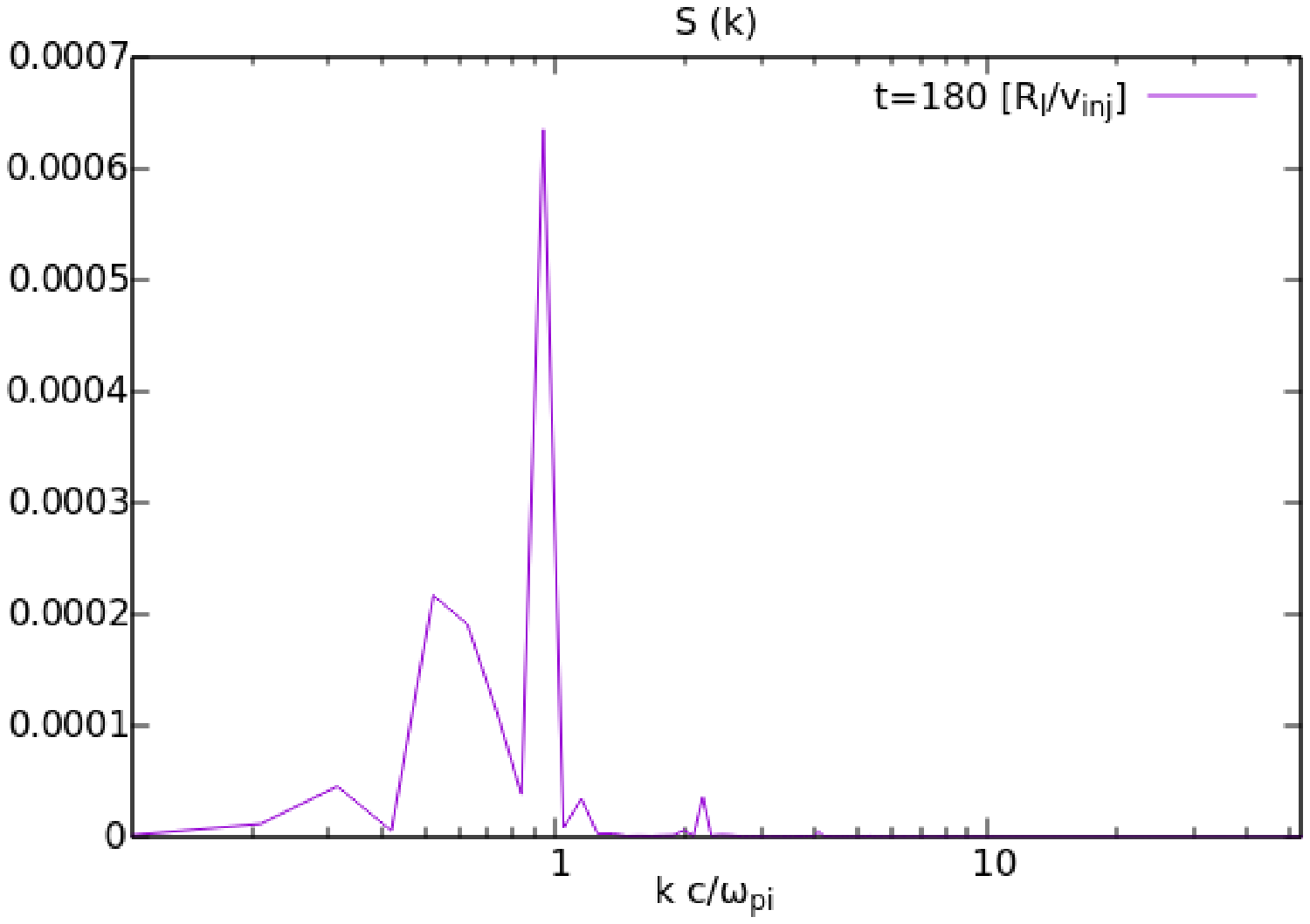}\\
\includegraphics[width=\columnwidth]{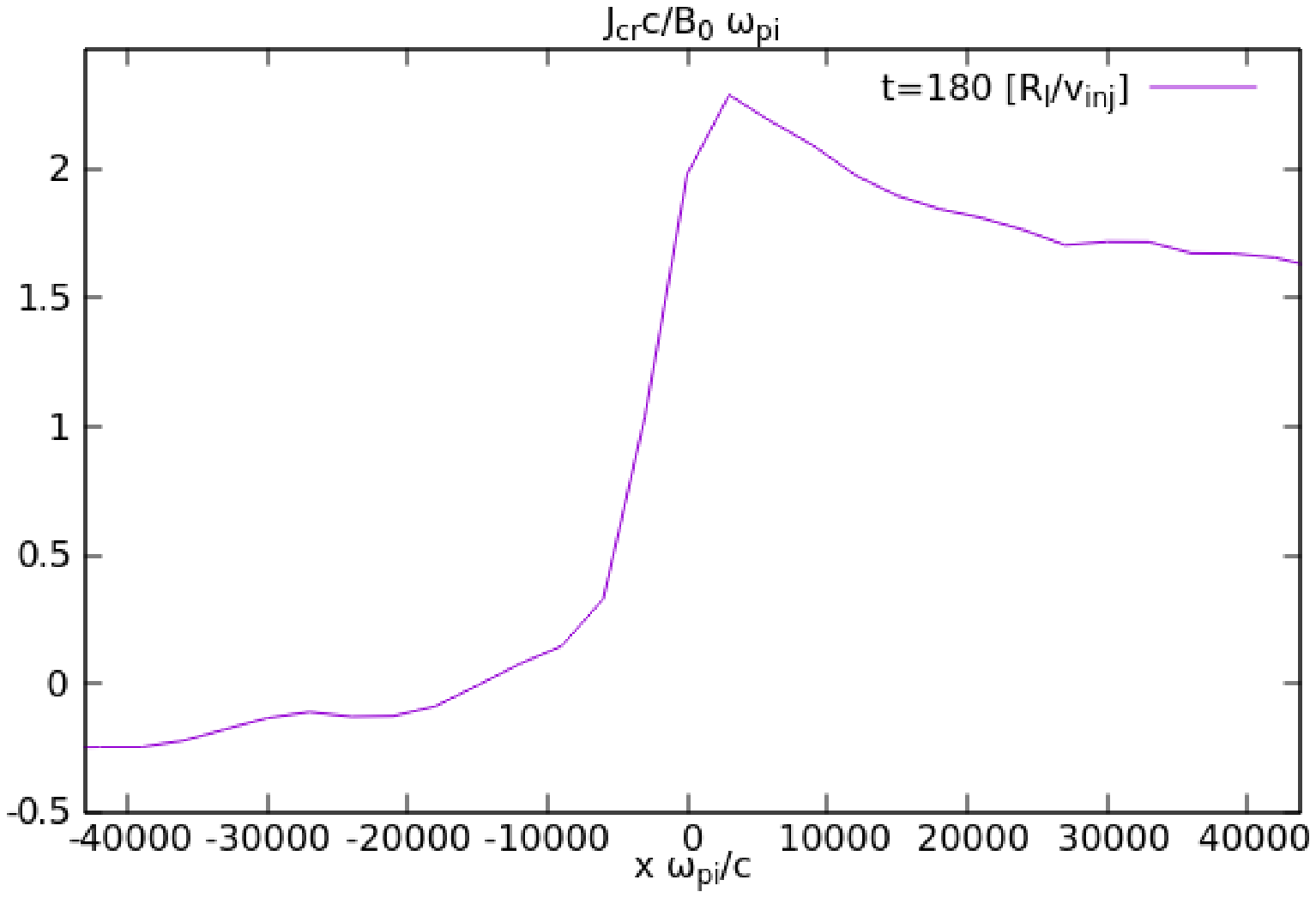}
\caption{The spectrum of the upstream transverse magnetic field (top) and the cosmic ray current (bottom) at $t=180\,R_l/v_{inj}$ (See Fig.~\ref{fig:picmhd1}.) We find a good agreement between the wave number of the dominant mode and upstream current.}
\label{fig:picmhdfft}
\end{figure}

According to the theoretical model by \citet{B04}, the fastest growing mode of the the non-resonant streaming instability would, in the linear phase, scale with the upstream current of the cosmic rays and the magnetic field as $k_{\rm max} =J_{\rm cr}/2B_0$. We check this by performing a FFT on the transverse variation of the upstream magnetic field near the shock at $t=180\,R_l/v_{inj}$. The result, expressed in the ion-cyclotron scale $c/{\omega_{\rm pi}}$, is shown in Fig.~\ref{fig:picmhdfft}. In the upstream medium directly ahead of the shock ($0<x\omega_{\rm pi}/c<20,000$), the cosmic ray current (bottom panel) is, on average, approximately 1.9. The dominant mode of the spectrum, at $k c/\omega_{pi}=0.9$, conforms to the theoretical prediction.

\begin{figure}
\includegraphics[width=\columnwidth]{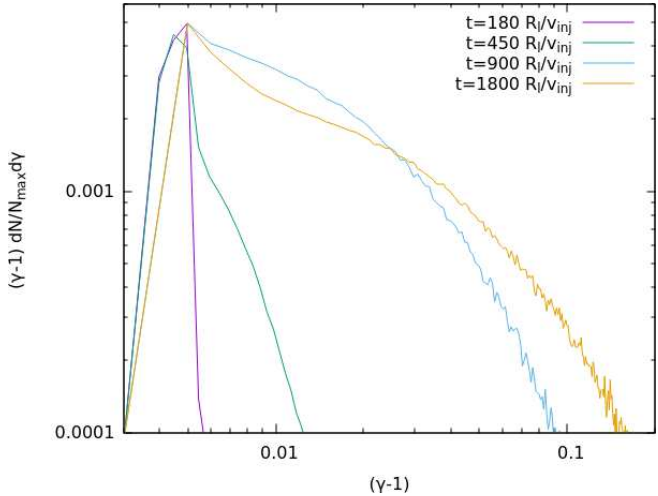}\\
\caption{Spectral energy distribution as a function of the Lorentz factor for the PIC-MHD simulation. From $t=450R_s/v_{\rm inj}$ onward the spectrum shows the distinctive high energy tail associated with DSA.}
\label{fig:picmhdsed}
\end{figure}

The disturbance of the magnetic field both upstream and downstream of the shock reflects the particles back toward the shock. This starts the DSA process. We hence find that in the conditions that prevail at these early shock expansion stages, efficient particle acceleration is occurring. This is proven by the spectral energy distribution (SED) of the particles in our simulation box, shown in
Fig.~\ref{fig:picmhdsed}. Initially, the particle distribution is centered at low energy. However, over time, a high energy tail appears (starting at  $t=450R_s/v_{\rm inj}$in Fig.~\ref{fig:picmhdsed}). N.B. because all particles in the PIC-MHD model are non-thermals, the SED does not show a Maxwellian distribution at low energy. Instead, the low-energy, thermal particles are represented by the MHD fluid.

\begin{figure}
\includegraphics[width=\columnwidth]{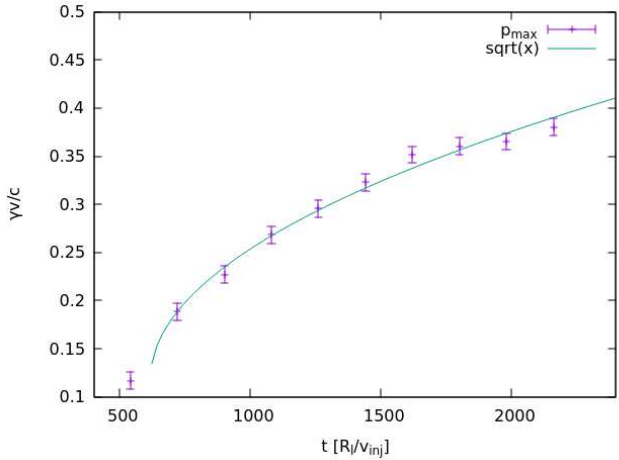}\\
\caption{Evolution of $p_{\rm max}$ as a function of time for the PIC-MHD simulation. Initially, the time evolution matches the predicted function well. In the later stages, it starts to deviate as the shock corrugation begins to dominate the shock structure.}
\label{fig:picmhdpmax}
\end{figure}

DSA theory predicts that the spectral distribution as a function of momentum should follow a power-law. This allows us to evaluate the maximum momentum by fitting the SEDs with the function, 
\begin{equation}
    f(p)~=~A_0 p^a \exp{(-p/p_{\rm max})}
\end{equation}
with $A_0$ a free parameter, $a$ the power-law index, and $p_{\rm max}$ the turn-off point where the SED starts to deviate from the DSA power-law. 
This gives us the evolution of $p_{\rm max}$ over time, which is plotted in Fig.~\ref{fig:picmhdpmax}. 
Theory predicts that $p_{\rm max}\,\propto\,\sqrt{t}$ if the diffusion coefficient is the Bohm one in the non-relativistic regime. The Bohm diffusion calculated using the total magnetic field is a good approximation in media where the perturbed component of the magnetic field exceeds the background one, see e.g. \citet{RB13}. As Fig.~\ref{fig:picmhdpmax} shows, the data points initially match the prediction rather well. However, after $t\,\gtrsim\, 1500$ they start to deviate, albeit still increasing. By this time, the shock has become severely corrugated, which influences the acceleration process, likely causing the deviation.

We can then advance that DSA is occurring very efficiently in such environments. As in the PIC-MHD simulations the maximum energy has not yet reached the relativistic regime we cannot at this stage estimate the time required to effectively reach 1 TeV, the injection energy in the kinetic-MHD runs described above.  But stating about an efficient magnetic field amplification at Gauss level, the TeV regime is reached in timescales $\ll$ day if the acceleration occurs in the Bohm regime for such fast shocks. This supports a fortiori the assumption made in section \ref{RES} to inject the particles at $p= 1$ TeV/c.

\subsection{Comparison with previous PIC and hybrid simulations}
Several previous (hybrid) PIC simulations reported magnetic field amplification up to the saturation level, while our simulations do not.
The particle-based simulations can study an evolution of CRs from an injection energy to an order of magnitude larger than the injection energy at best.
Given that the saturation level of the NRH instability depends on the maximum energy of CRs (see, eq.~[\ref{eq:Bsat}]), the small available maximum energy in the particle based simulation makes it easy for the NRH instability to saturate.
\citet{CS14b}) showed, in their seminal paper, that magnetic energy amplification factor $\langle B^2/B^2_{\rm ini}\rangle$ increase linearly with $M_{\rm A}$ as $\langle B^2/B^2_{\rm ini}\rangle \simeq 0.5\,M_{\rm A}$ in the range $10\le M_{\rm A}\le 100$.
The result of our simulations suggests that the relation possibly does not hold for Mach numbers $M_{\rm A}\gtrsim 10^3$. A possible explanation could be that when the shock speed becomes a non negligible fraction of the light speed advection effects start to compete with the instability growth and may prevent for reaching the expected saturation level. Note that if we compare the amplification factor itself, the results of our simulation shows larger values than that of the PIC simulations (e.g., $\langle B^2/B^2_{\rm ini}\rangle\sim 100$ for Model 0). 

\subsection{Effect of CR pressure}\label{pcr}
In the present simulations, we did not take into account the effect of CR pressure that modifies shock structure making CR precursor and influences the CR spectrum, depending on the CR injection rate \citep{M97a, M97b, BGV, VYH10, SHA13}.
Semi-analytic studies of steady state CR modified shock revealed that there are three possible states of the CR modified shock when the CR injection rate is set to be in a certain range $\eta_2\ge\eta\ge\eta_1$. The so-called efficient state has large compression ratio and substantially modifies CR spectrum, while the inefficient state has smaller influence on the shock structure. Finally, there is an intermediate state in between the efficient and inefficient solutions. 
In the case $\eta\ge\eta_2$ ($\eta\le\eta_1$), there is only the efficient (inefficient) solution. Time-dependent analysis by \citet{SHA13} has shown that, for $\eta\le\eta_2$, the inefficient solution would be naturally selected, because the efficient and inefficient solutions are stable against perturbations, while the intermediate one is unstable.
Thus, so far as $\eta\le\eta_2$, the neglect of the effect of CR pressure would be reasonable as a first approach. According to \citet{M97b}, for $p_{\rm inj}\,c= 6\times 10^7$ eV and $p_{\rm max}\,c= 10^{15}$ eV that are compatible with our simulations, the critical injection rate is estimated to be $\eta_2\simeq 10^{-3}$, which is roughly twice larger than our fiducial choice of $\eta$ (it should be noted that the definition of $\eta$ is different between this paper and \citet{M97b}, and we have adapted to our definition).

However, even in the inefficient case, the CR precursor has still non-negligible influences on the shock structure and resulting CR spectrum.

In our injection rate model given by eq.~(\ref{Qinj}), we assume the standard DSA spectrum for $p\le p_{\rm L}$. This assumption is reasonable because the injection is modest in all our models and we do not anticipate to have strong shock modification due to CR pressure effects. But if we consider the CR pressure modified shock, the CR spectrum with $p\lesssim 1$ GeV $c^{-1}$ can be steeper than that of standard DSA depending on the injection rate. One may have a such an effect if we consider another strong feed back produced by Alvf\'enic drifts \citep{Diesing21}. However, as the fraction of the shock kinetic energy imparted in CRs or magnetic fluctuations does not exceed a few \%, we do not expect a CR solution to be strongly different form $p^{-4}$. A softer spectrum potentially weakens the CR current and growth of the NRH instability \citep{Araudo21} even if a dedicated investigation is necessary to validate this assertion. In the present spatial resolution, it is hard to resolve the detailed structure of the CR modified shock, which requires resolving the diffusion length for 1 GeV CRs. The effect of CR modified shock can be integrated into the present model, including drift effects, once we find an appropriate interpolation of the CR spectrum below $p_{\rm L}$, which we leave as our future works.

\subsection{Restrictions due to 1D geometry}
Since the NRH instability is essentially 1D phenomenon, the growth of which does not seem to depend on spatial dimensionality even in non-linear phase.
However, if we extend our system to that involves the effects of cosmic-ray pressure and multi dimensionality, we can expect additional magnetic field amplification by the Drury instability mediate turbulent dynamo \citep{BJL09}.
For the upstream, this effect can amplify magnetic field in the precursor region of the CR modified shock, which helps to enlarge the maximum energy of the CR. We plan to include pressure and multi dimensionality effects in our future code developments.

\subsection{Implications for observations}
The onset time of the collisionless shock (CS) during the early SN envelope expansion phases determines the start of particle acceleration and multi-wavelength radiation \citep{Levinson20}. The time after which the CS forms depends on the properties of the upper stellar atmosphere or circum-stellar wind.
\cite{GB15} give a constrain of the shock speed for the CS to form before shock breakout. Namely $v_{\rm sh} \lesssim 0.1\,c \times (v_{\rm w}/ 10~\rm{km s}^{-1})$ $(\dot{M}/5\times10^{-4}~M_\odot\,\rm{yr}^{-1})^{-1} (r_{\star}/10^{13}~\rm{cm})$, where $v_{\rm w}$, $\dot{M}$ and $r_{\star}$ are the wind speed, mass loss rate (in solar mass per year) and progenitor core radius respectively. In that aspect, RSG winds with enhanced mass loss prior to the explosion can fulfill this constrain. If the scenario can operate in core collapse SNe, then primary or secondary (those particles produced by primary CR interaction with matter and/or radiation) CRs distribution can be traced through their multi-wavelength radiation. A sub-sample of core collapse SNe show non-thermal radio emission associated to synchrotron radiation from mildly relativistic or relativistic electrons as early as a few days after the SN flash. This was in particular the case for the nearby (at a distance of $\sim$ 3.5 Mpc) type IIb SN 1993J which showed non-thermal radio emission 5 days after the outburst \citep{Weiler07}. But the main channels allowing a probe of the acceleration of high energy hadrons either are  the X-ray emission due to secondary leptons and the gamma-ray emission mostly due to neutral pion decay from the interaction of high energy CRs with the CSM wind material \citep{Marcowith14}. Gamma-ray photons can be either produced through Inverse Compton emission by relativistic electrons, but at least to what concerns energies above 10 TeV, the Klein-Nishina effect should limit the lepton contribution to the gamma-ray emission strongly. \cite{Dwarkadas13} calculate the expected level of hadronic gamma-ray emission for a large variety on ejecta and density profile in the CSM but did not consider gamma-gamma pair production attenuation. \cite{Murase19} introduce the effect of pair production and the electromagnetic cascade initiated by secondary particles issued from CR interaction with the surrounding material. They propose a parametric one-zone model for different classes of core-collapse supernovae. \cite{T09} investigate the gamma-ray emission from SN 1993J more specifically. To that aim he developed a model based on microphysics of shock acceleration which main parameters (for instance the CR injection fraction) are constrained to reproduce radio lightcurves from a few days after the outburst. He found a flux above 1 GeV of $F \simeq 2\times10^{-9}$ $ (t/1\,\rm{day})^{-1}$ cm$^{-2}$ s$^{-1}$ weakly subject to gamma-gamma attenuation and an unabsorbed flux above 1 TeV $F \simeq 2\times10^{-12}$ $(t/1\,\rm{day})^{-1}$ cm$^{-2}$ s$^{-1}$.
But at these energies during the first 10 days after the outburst the effective flux is well below this limit due to gamma-gamma attenuation.
In this latter model gamma-gamma absorption is calculated considering an isotropic flux of soft photons. However, as shown by \cite{CRM20} gamma-gamma attenuation involves in fact an anisotropic and a time-dependent flux of soft photons which render the detailed calculation much more complex.
They applied their calculation to the case of SN~1993J and showed that the source could be barely detected by the Cerenkov Telescope Array (CTA).
However, in their calculation they considered a constant mass loss rate of $\dot{M} \simeq 3\times10^{-6} M_{\odot}\,\rm{yr}^{-1}$.
A strong mass loss enhancement in the few years before the explosion should greatly improved the gamma-ray detectability of this category of source by upcoming gamma-ray facilities such as the CTA.
Let us mention here the potential interest for this type of study of other ground-based facilities like the Tibet AS-gamma experiment, the High-Altitude Water Cherenkov (HAWC) observatory, the Large High Altitude Air Shower Observatory (LHAASO), and the Southern Wide-field Gamma-ray Observatory (SWGO).
These experiments still have a good sensitivity above 10 TeV so they should be able to more strongly probe any PeVatron hadronic signature in these objects.
Besides, as gamma-gamma attenuation is expected to be strong at least within the first week after the explosion \citep{CRM20}, an important work is mandatory to coordinate the multi-wavelength follow-up campaigns including gamma-ray instruments.
Finally, as CRs will interact with CSM they can also produce charged pions and hence high energy neutrinos. A preliminary calculation in the case of SN 1993 J show that km$^3$ neutrino telescopes should have barely detected the source \citep{Marcowith14}, but again the calculations were conducted there assuming a relatively low stellar mass loss rate prior to the explosion. Enhanced mass loss or other type of core-collapse supernovae may be sources of high energy neutrinos \citep{Zira16}.

\section{summary}
Using a hybrid code that solves the Bell MHD equations and the telegrapher-type diffusion convection equation, we have performed simulations of cosmic-ray acceleration at a supernova blast wave shock propagating in dense CSM created by RSG wind.
We have found that, under the fiducial parameter setting, the NRH instability can amplify magnetic fields actively and helps to enhance maximum cut-off energy of cosmic-rays up to approximately 1 PeV in $\sim10$ days after the explosion, although the amplified magnetic field does not reach the saturation level owing to geometrical and finite spatial extension of escaped CRs (see \S \ref{sec:model0}).
When we employ an enhanced mass loss rate of the RSG wind ($\dot{M}=10^{-3}$ M$_{\sun}$ yr$^{-1}$), which is pointed out to be more realistic by recent SNe observations \citep{FMM18}, the maximum cut-off energy can be as large as the knee energy (3 PeV).
Thanks to the enhanced mass-loss rate, such SNe could be visible as PeVatrons through the  upcoming  gamma-ray facilities.

\acknowledgments
The numerical computations were carried out using the supercomputer "Flow" at Information Technology Center, Nagoya University and XC50 system at the Center for Computational Astrophysics (CfCA) of National Astronomical Observatory of Japan.
This work is supported by Grant-in-aids from the Ministry of Education, Culture, Sports, Science, and Technology (MEXT) of Japan (20H01944). This work is supported by the ANR-19-CE31-0014 GAMALO project.

\begin{figure}[h]
\begin{center}
\includegraphics[scale=0.7]{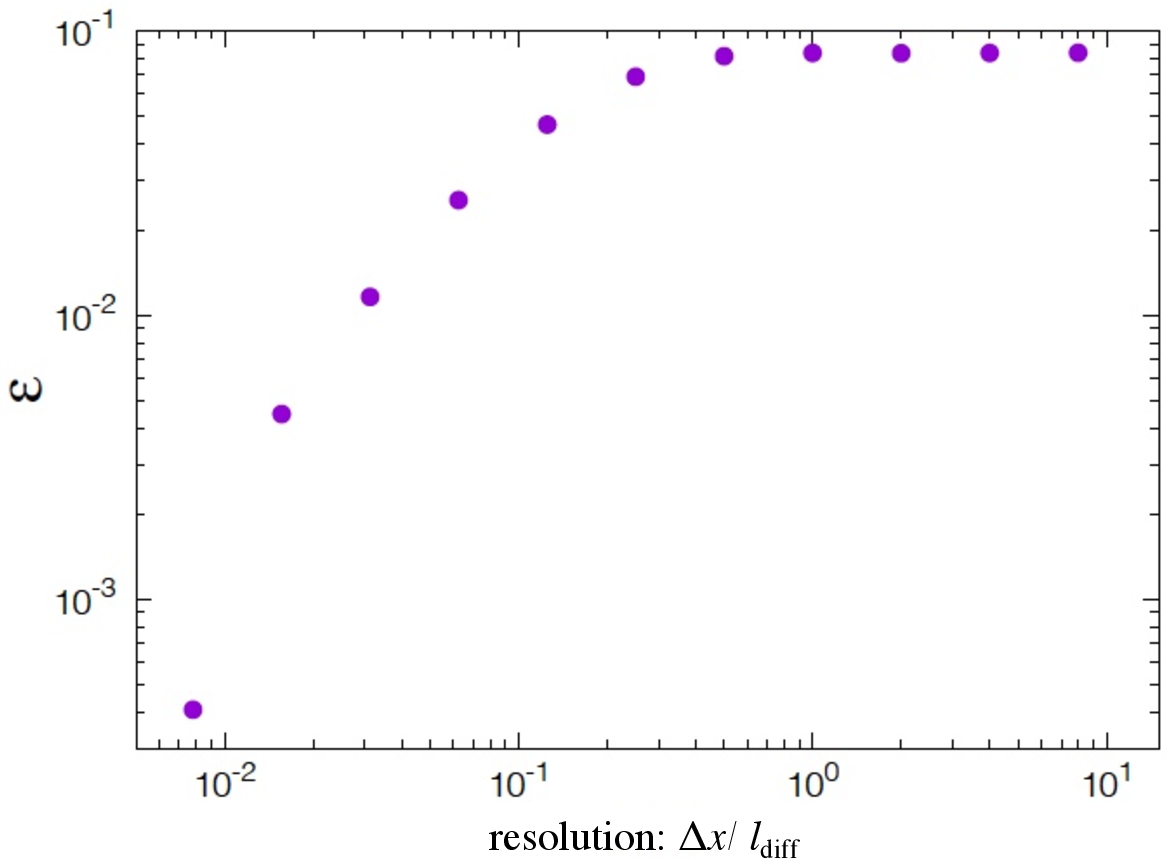}
\caption{
CR spectral index deviation $\epsilon$ from the standard DSA value ($p=3\,r_{\rm c}/(r_{\rm c}-1)$) as a function of the spatial resolution (spatial cell width divided by the CR diffusion length).
The back ground shock wave has Mach number $M_{\rm s}=100.0$ ($r_{\rm c}=3.9988$), and test runs are performed in the shock rest frame.
}
\label{fApp1}
\end{center}
\end{figure}

\appendix
In this section, we show the results of basic tests that indicates ability of our code.
Figure \ref{fApp1} is the plot of the numerically obtained CR spectral index deviation $\epsilon$ from the well known analytic DSA value ($3\,r_{\rm c}/(r_{\rm c}-1)$) as a function of the spatial resolution, where $r_{\rm c}$ is the compression ratio of the shock.
For the test runs, we set a plane parallel, hydrodynamic shock of Mach number $M_{\rm s}=100.0$ ($r_{\rm c}=3.9988$) in the shock rest frame, and set a constant diffusion coefficient for CRs.
The numerical spectral index is obtained by fitting CR spectrum $f_0$ after reaching a steady state.
We see that the spectral index error $\epsilon$ is less than $0.01$ ($0.25\%$) once the diffusion length of the CRs ($l_{\rm diff}=\kappa/v_{\rm sh}$) are resolved more than $\sim 10$ cells, which is satisfied in our simulations.
Such a low error cannot be obtained if the shock is not captured finely with appropriate compression ratio.

\begin{figure}[h]
\begin{center}
\includegraphics[scale=0.8]{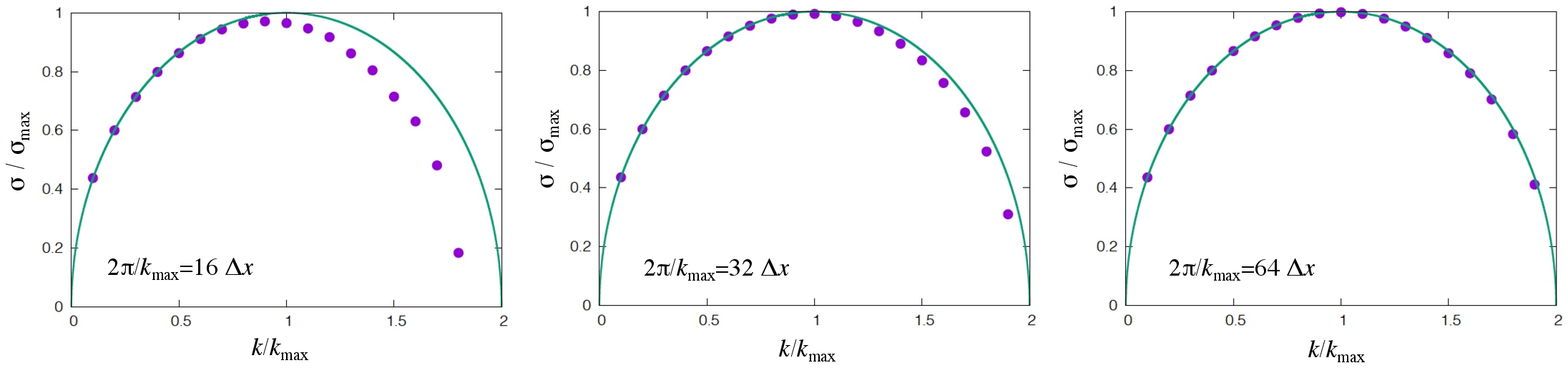}
\caption{
Result of test simulations of the Bell instability growth in linear phase (points).
Solid lines are the theoretical dispersion relation under a given CR current.
Horizontal and vertical axes are normalized, respectively, by the most unstable wave number ($k_{\rm max}=2\pi\,j^2/B\,c$) and by the growth rate $\sigma$ at $k_{\rm max}$.
From left panel to right panel, the spatial resolution ($\lambda_{\rm max}/\Delta x$) is increased from 16 to 64.
}
\label{fApp2}
\end{center}
\end{figure}

Figure \ref{fApp2} is the comparison between numerically measured growth rate of the NRH instability and analytic solution under a given CR current ($\sigma={\rm Im}[ \{v_{\rm A}^2\,k^2-B\,j^{(\rm{ret})}\,k/\rho\,c\}^{1/2}]$).
We can confirm that our scheme reproduces the growth rate of the NRH instability, once the unstable scale is resolved more than $\sim$ 10 cells, which is satisfied as discussed in \S 2.4.

\end{document}